\documentclass[prb,amsmath,reprint,showpacs]{revtex4-1}
\usepackage{graphicx}
\usepackage[version=3]{mhchem}
\DeclareMathOperator{\trace}{Tr}
\DeclareMathOperator{\sign}{sgn}
\providecommand{\abs}[1]{\lvert#1\rvert}
\providecommand*{\unit}[1]{\ensuremath{\mathrm{\,#1}}}

\begin{document}
\title{Dysprosium-based experimental representatives of an Ising-Heisenberg chain
\\and a decorated Ising ring}
\date{\today}
\author{Willem \surname{Van den Heuvel}}
\email{willem.vandenheuvel@chem.kuleuven.be}
\author{Liviu F. Chibotaru}
\email{liviu.chibotaru@chem.kuleuven.be}
\affiliation{Chemistry Department, Katholieke Universiteit Leuven,
Celestijnenlaan 200F, B-3001 Leuven, Belgium}

\begin{abstract} It is shown that the bond-decorated Ising
model  is a realistic model for certain real magnetic compounds containing
lanthanide ions.  The lanthanide ion plays the role of Ising spin. The required
conditions on the crystal-field spectrum of the lanthanide ion for the model to
be valid are discussed and found to be in agreement with several recent
\textit{ab initio} calculations on \ce{Dy^{3+}} centers. Similarities and
differences between the spectra of the simple Ising chain and the decorated
Ising chain are discussed and illustrated, with attention to level crossings in
a magnetic field. The magnetic properties of two actual examples (a
[DyCuMoCu]$_\infty$ chain and a \ce{Dy4Cr4} ring) are obtained by a
transfer-matrix solution of the decorated Ising model.  $g$-factors of the
metal ions are directly imported from \textit{ab initio} results, while
exchange coupling constants are fitted to experiment. Agreement with experiment
is found to be satisfactory, provided one includes a correction (from
\textit{ab initio} results) for susceptibility and magnetization to account for
the presence of excited Kramers doublets on \ce{Dy^{3+}}.  
\end{abstract} 

\maketitle

\section{Introduction}\label{4:sec:intro}

In a 1959 paper M.~E.~Fisher introduced the general bond-decorated Ising model
as one example of a set of exactly solvable transformations of spin-1/2 Ising
models.\cite{PR_113_969} A bond-decorated Ising model has an ``arbitrary
statistical mechanical system'' inserted in every original Ising bond. The
partition function of this decorated model is related to the partition function
of the original or bare Ising model by the addition of a prefactor and a
renormalization of the coupling constants and magnetic moments (the Ising model
is supposed to be in a parallel magnetic field).\cite{PhysLettA_374_3718}
Knowledge of the partition function of a given Ising model thus allows one to
obtain the partition function of any bond-decorated version of that Ising
model.

\begin{figure}
\centering
\includegraphics{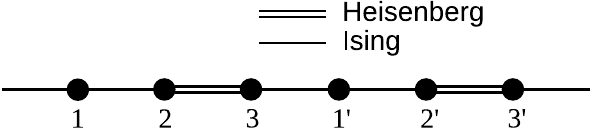}
\caption{\label{4:fig:IsingHeisenberg}The Ising-Heisenberg chain discussed in
Ref.~\onlinecite{CzechJPhys_52_A37} (showing only two unit cells). The bonds of
the Ising chain (1-$1'$-$1''$- \ldots)\ are decorated with Heisenberg dimers 
$\big($(2-3), ($2'$-$3'$), \ldots$\big)$. The partition function of this chain is exactly solvable.}
\end{figure}

More recently, Stre\v{c}ka and Ja\v{s}\v{c}ur have used the method of
bond-decoration to investigate the thermodynamics of mixed Ising-Heisenberg
chains in parallel magnetic fields where the decorating unit is a spin dimer or
trimer with anisotropic Heisenberg coupling (see
Fig.~\ref{4:fig:IsingHeisenberg}).\cite{CzechJPhys_52_A37,PRB_72_024459} The
partition function of these chains is readily obtained from the known partition
function of the Ising chain and the energy levels of the decorating unit.
Therefore, they could calculate exact magnetic properties and theoretically
show, for example, the existence of magnetization plateaus in certain
bond-alternating chains.

The convenience of the decorated Ising chain as a theoretical model for spin
chains derives from the relative ease with which exact solutions are obtained,
in contrast for example with the pure Heisenberg chain, for which no exact
partition function has been found. Up to now, this property of solvability has
been the prime motive for the study of these chains in the literature. Indeed,
in Ref.~\onlinecite{CzechJPhys_52_A37} the decorated Ising chain was considered
as a substitute for the intractable Heisenberg model, and in
Ref.~\onlinecite{PRB_72_024459} the principle reason for introducing Ising
bonds---to replace the more reasonable Heisenberg bonds---in a chain of
Cu\textsuperscript{2+} ions was the desire to obtain a solvable model. This
approach can be applied to any type of Heisenberg chain with a repeating unit:
replace enough---preferably ferromagnetic---Heisenberg bonds by Ising bonds to
obtain a decorated Ising chain that is solved easily and, in some cases,
exhibits thermodynamic properties that are qualitatively comparable with those
of the original chain.\cite{PRB_79_014432,CondensMatterPhys_12_343}

However, the role of the decorated Ising chain in the field of one-dimensional
magnetism is not confined to that of a simplified model of realistic quantum
spin chains. In this paper we show that some new molecular rings and chains are
real examples of decorated Ising systems. Concretely, we treat a [DyCuMoCu]
infinite chain\cite{ChemEurJ_15_11808} and a (DyCr)$_4$ tetrameric
ring.\cite{AngewChemIntEd_49_7583} These compounds were recently synthesized
in the course of the ongoing synthetic efforts to make new and better
single-chain magnets (SCMs) and single-molecule magnets (SMMs), whose
characteristic property is a blocking or slow relaxation of magnetization at
low temperatures.\cite{[{For single molecule magnetism see
}]Gatteschi_mol_nanomagn,*[{for single chain magnetism see }]SCM_coulon} We
will not be concerned here with these dynamical aspects of their magnetism, but
only with their static magnetic properties. A necessary property of SMMs and
SCMs is a magnetic anisotropy. One line of approach is to introduce anisotropy
by means of lanthanide ions, whether or not in combination with transition
metal ions.\cite{InorgChem_48_3342,CoordChemRev_253_2009} The two compounds
considered here are products of this approach, with dysprosium as lanthanide
ion.

The Dy$^{3+}$ ion plays a crucial role in these systems; the nature of the
ground state of this ion in its ligand environment determines whether the
system is a decorated Ising system or not and consequently, whether its
partition function is exactly solvable or not. The ground Kramers doublet of
Dy$^{3+}$ must have complete uniaxial magnetic anisotropy\footnote{By complete
uniaxial anisotropy we mean that only one $g$-factor of the Kramers doublet is
not zero; for example $g_x=g_y=0$ and $g_z\neq 0$.} and must be separated from
excited Kramers doublets by an amount that is large compared with the exchange
coupling (typically, this separation must be 100~cm$^{-1}$ or
more).\cite{BrazJPhys_30_794} The required information on the ground and
excited doublets of the Dy$^{3+}$ monomer can be derived from \textit{ab
initio} calculations on the monomer complex, isolated from the polynuclear
compound.\cite{ChemEurJ_15_11808,AngewChemIntEd_49_7583}

The [DyCuMoCu] chain and the (DyCr)$_4$ ring are shown to be decorated Ising
chains in an arbitrarily directed magnetic field. The magnetic properties, in
particular powder magnetization and susceptibility, are calculated with the
help of the transfer-matrix method, which is a bit more general and convenient
for numerical computation than the renormalization of the Ising parameters,
which was used by Fisher and by Stre\v{c}ka and Ja\v{s}\v{c}ur. The results
compare well with experiment and allow to determine values for the
exchange coupling constants. 

The excited crystal field Kramers doublets (or Stark levels) of Dy$^{3+}$ are
not included in the decorated Ising model. Because of their relatively low
energetic position, these Kramers doublets can have a non-negligible
contribution to the magnetic properties of the chain. We account for this in a
first approximation by adding this contribution, as calculated \textit{ab
initio} for the monomeric Dy$^{3+}$ complex, to the results of the decorated
Ising model.

\section{Theoretical Background}\label{4:sec:theory}
\subsection{Decorated Ising chain and transfer-matrix solution}
\label{4:sec:decIsing}
\begin{figure}
\centering
\includegraphics[width=8.6cm]{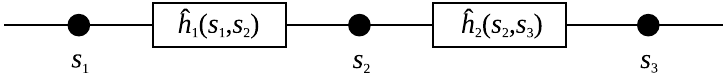}
\caption{\label{4:fig:decIsingchain}Part of the decorated Ising chain. In each
Ising bond is inserted an arbitrary statistical mechanical
system\cite{PR_113_969}, called decorating unit, that interacts only with the
two Ising spins $s_i$ at the vertices of the bond. The Ising spin variables
commute with the Hamiltonian and have definite values in the eigenstates of the
chain.} \end{figure}

The decorated Ising chain may be divided into units delimited by the Ising
spins, as in Fig.~\ref{4:fig:decIsingchain}. The Hamiltonian of the decorated
Ising chain (or ring) of length $n$ may accordingly be written as follows:
\begin{equation}\label{4:eq:HamdecIsing}
\hat{H}=\hat{h}_1(s_1,s_2)+\hat{h}_2(s_2,s_3)+\hat{h}_3(s_3,s_4)+\ldots+
\hat{h}_n(s_n,s_{n+1}),
\end{equation}
where the subhamiltonians $\hat{h}_i$ correspond to units of the chain. If the
chain is closed into a ring, periodic boundary conditions apply by identifying
the last with the first spin: $s_{n+1}\equiv s_1$. The Ising spins $s_i$
commute with the subhamiltonians and the subhamiltonians commute with each
other:
\begin{equation}\label{4:eq:commutation}
[s_i,\hat{h}_j]=0 \quad \text{and} \quad [\hat{h}_i,\hat{h}_j]=0.
\end{equation}
It is further assumed that there is no direct interaction between the
decorating units themselves, i.e., two different $\hat{h}_i$ have no variables
in common, except for an $s_i$ when they are neighbors. In writing
Eq.~(\ref{4:eq:HamdecIsing}) we made use of the fact that the Ising spins are
conserved variables and may be considered as parameters rather than operators.
In the case of spin-1/2, they take on the values $+1/2$ and $-1/2$, so that to
each decorating unit there correspond four different subhamiltonians
$\hat{h}_i(s_i,s_{i+1})$.\footnote{In what follows, no use is made of the
angular momentum properties of the $s_i$. In fact, any parameter that takes on
a finite number of values can serve as $s_i$ in Eq.~(\ref{4:eq:HamdecIsing}).
In practice however, $s_i$ represents most commonly a Kramers doublet, which is
most often described as an effective spin-1/2. This is the case for the
examples considered in this paper.} 

Since there is no direct interaction between the decorating units of the chain,
the subhamiltonians $\hat{h}_i$ in Eq.~(\ref{4:eq:HamdecIsing}) work---for a
given set of $s_i$ values---on disjunct spaces. An eigenfunction of $\hat{H}$
is then simply a direct product of $n$ independent eigenfunctions, one of each
$\hat{h}_i$.  The corresponding energy is:
\begin{equation}\label{4:eq:energies}
\begin{split}
E&=\varepsilon_{1k_1}(s_1,s_2)+\varepsilon_{2k_2}(s_2,s_3)+\varepsilon_{3k_3}
(s_3,s_4)+\ldots\\
&\quad +\varepsilon_{nk_n}(s_n,s_{n+1}),
\end{split}
\end{equation}
where $\varepsilon_{ik_i}(s_i,s_{i+1})$ is the $k_i$-th energy level of
$\hat{h}_i(s_i,s_{i+1})$.

Following Ref.~\onlinecite{PR_113_969} we write the partition function of one
decorating unit for fixed $s_i$ and $s_{i+1}$ on the bond vertices:
\begin{equation}\label{4:eq:subpartitionfunction}
\psi_i(s_i,s_{i+1})=\sum_k \exp[-\beta\varepsilon_{ik}(s_i,s_{i+1})],
\end{equation}
where $\beta=1/kT$. The total partition function for the chain is then given by
\begin{equation}\label{4:eq:partitionfunction}
\mathcal{Z}_n=\sum_{s_1,s_2,\ldots,s_{n+1}}
\psi_1(s_1,s_2)\psi_2(s_2,s_3)\ldots
\psi_n(s_n,s_{n+1}).
\end{equation}
This is the well-known form of a transfer-matrix solution.\cite{Yeomans}
Each decorated bond (or pair of neighboring Ising spins) is represented by a
transfer matrix $T_i$ whose elements are the $\psi_i(s_i,s_{i+1})$, the values
of $s_i$ labeling the rows and the values of $s_{i+1}$ labeling the columns.
Explicitly for the spin-1/2 decorated Ising chain:
\begin{equation}\label{4:eq:Tmatrix}
T_i=\begin{pmatrix} \psi_i(+\frac{1}{2},+\frac{1}{2}) &
\psi_i(+\frac{1}{2},-\frac{1}{2})\\ \psi_i(-\frac{1}{2},+\frac{1}{2})&
\psi_i(-\frac{1}{2},-\frac{1}{2}) \end{pmatrix}.
\end{equation}
Since most applications deal either with rings or very long chains, the
boundaries can be identified ($s_{n+1}\equiv s_1$) and expression
(\ref{4:eq:partitionfunction}) can be written as the trace of a matrix product:
\begin{equation}\label{4:eq:partfunc_transmatr}
\mathcal{Z}_n=\trace(T_1T_2\ldots T_n).
\end{equation}
This is the most general expression for the partition function of a decorated
Ising chain. If the chain has translational symmetry the partition function is
expressed in terms of eigenvalues.\cite{Yeomans} Suppose that the chain is a
repetition of identical decorated Ising bonds, then the $T_i$ are all the same,
so that
\begin{equation}\label{4:eq:partfunc_periodic}
\mathcal{Z}_n=\trace(T_1^n)=\sum_i\lambda_i^n,
\end{equation}
where the $\lambda_i$ are the eigenvalues of $T_1$. According to Perron's
theorem on positive matrices, the eigenvalue $\lambda_0$ with largest modulus
is real, positive and nondegenerate.\cite{Horn_Johnson_matrixanalysis} This
yields the simple but exact result for the free energy per unit cell of the
infinite chain:
\begin{equation}\label{4:eq:freenergy_infinite}
f=-kT\lim_{n\to\infty}\frac{\ln\mathcal{Z}_n}{n}=-kT\ln\lambda_0.
\end{equation}
If a unit cell of the chain is spanned by $p$ decorated bonds instead of one
then it is only necessary to combine $p$ transfer matrices into one new
transfer matrix $\widetilde{T}=T_1T_2\ldots T_p$, with largest eigenvalue
$\tilde{\lambda}_0$, to be used in Eqs.~(\ref{4:eq:partfunc_periodic}) and
(\ref{4:eq:freenergy_infinite}), where $n$ is to be replaced by $n/p$. This
situation arises, for example, when the local easy axes on the Ising ions are
not parallel but canted with respect to each other.

The solution in terms of transfer matrices,
Eq.~(\ref{4:eq:partfunc_transmatr}), is not limited to decorated spin-1/2 Ising
chains. It is valid for chains having Ising spins of any multiplicity or a
combination of Ising spins with different multiplicities, in which case the
dimension of the transfer matrix is different from two by two.  Another
advantage of the transfer matrix method is that it can be readily extended to
include next-nearest-neighbor bonds between the Ising spins. To this end, the
transfer matrix has to be enlarged so that it does not jump from one Ising spin
to the next, but from one \emph{pair} of Ising spins to the next
pair.\cite{Yeomans}

\subsection{\ce{Dy^3+} as Ising spin}\label{4:sec:Dy}

We are interested in this paper in molecular chains or rings that can be
described by a decorated Ising model. This means that their low-energy spectrum
can be modeled to satisfactory accuracy by an effective Hamiltonian that has
the properties described in Section \ref{4:sec:decIsing}. It has been noted
there that the composition of the decorating unit is basically arbitrary and we
need therefore not consider the properties of that part. Instead, our attention
goes here to the molecular realization of the Ising spin [$s_i$ in
Eq.~\eqref{4:eq:HamdecIsing}]. We focus on \ce{Dy^3+} because this ion is used
in the two examples studied below, but the discussion applies equally well to
several other trivalent lanthanides and also to some transition metal ions.

It is known that lanthanide ions in a coordination environment are often well
described by crystal field theory applied to the ground $|LSJ\rangle$ level.
One assumes that $J$, $L$, and $S$ remain good quantum numbers. \ce{Dy^3+} is a
Kramers ion that belongs to the second half of the lanthanide series, whose
ground level is \ce{^6H_15/2} ($f^9$), with associated Land\'e factor $g=4/3$.
This multiplet splits into eight Kramers doublets by the crystal field
perturbation (except for high-symmetric environments belonging to the $T$, $O$,
or $K$ point groups, which split the multiplet in less than eight levels). It
will be useful to view each Kramers doublet as an effective spin-1/2 with its
own $g$-factors (3 in number) and corresponding magnetic axes. For example,
take the Kramers doublet $|M_J=\pm15/2\rangle$ (of the \ce{^6H_15/2} level),
quantized with respect to the $z$ axis. Its $g$ factors are
$g_z=2(4/3)(15/2)=20$ and $g_x=g_y=0$. 

If the action of the crystal field on \ce{Dy^3+} is such that the lowest
Kramers doublet is separated from the next one by an energy that is large
compared to the energy of interaction with the magnetic field and the exchange
interaction with neighboring ions, we can omit all excited Kramers doublets
from the Hamiltonian and keep only the lowest doublet. In this way \ce{Dy^3+}
is described by a spin of 1/2 and every interaction in which it takes part
enters the Hamiltonian as a linear combination of the three spin operators
$s_x$, $s_y$, and $s_z$.\footnote{Linear combination in a general sense: the
coefficients of the expansion may be operators working in other spaces.} If we
now want \ce{Dy^3+} to be an Ising spin, defined by the first commutation
relation in Eq.~\eqref{4:eq:commutation}, it is clear that only one of these
spin components, say $s_z$ or simply $s$, may actually appear in the
Hamiltonian. In other words, there must be no interaction that creates an
off-diagonal matrix element between the two components of the Kramers doublet.
This can be shown to be true with high accuracy if the lowest Kramers doublet
is $|M_J=\pm15/2\rangle$. 

The two interactions of importance here are the Zeeman interaction with the
magnetic field and the exchange interaction with other magnetic ions. The
Zeeman Hamiltonian follows directly from the $g$ factors of the Kramers doublet
(vide supra), and is 
\begin{equation}\label{4:eq:zeeman}
-\mu_\mathrm{B}g_zs_zB_z, 
\end{equation} 
where $B_z$ is the $z$ component
of the applied magnetic field. Note that the field may be applied in any
direction, but it is only the $z$ component that interacts with the Kramers
doublet because $g_x=g_y=0$. The vanishing of $g_x$ and $g_y$ in
$|M_J=\pm15/2\rangle$ follows from the selection rule stating that a vector
operator (the magnetic moment in this case) cannot connect states for which
$M_J$ differs by more than one unit. We will sometimes refer to the $z$ axis as
the anisotropy axis, to stress that it is the only magnetic axis with
nonvanishing $g$ factor. 

To evaluate the effect of exchange interaction, we must first take a closer
look at the composition of the Kramers doublet. In terms of the Russel-Saunders
states $|M_S\rangle|M_L\rangle$ we have
\begin{equation}\label{4:eq:KD}
\begin{split}
|+15/2\rangle &= |+5/2\rangle|+5\rangle\\
|-15/2\rangle &= |-5/2\rangle|-5\rangle.
\end{split}
\end{equation}
In a basic (super)exchange process between two magnetic centers, one electron
of each center takes part. If we look at one center, the process removes an
electron with certain spin projection (up or down) and puts it back on the
center either with the same or with opposite spin
projection.\cite{PhYsRev_115_2} This gives rise to the selection rule $\Delta
M_S=0,\,\pm 1$. If the exchange interaction is to connect both components of
the Kramers doublet in Eq.~\eqref{4:eq:KD}, at least five successive processes
are needed, for $\Delta M_S=\pm 5$. In other words, not one but five electron
spins have to be flipped to connect $|+15/2\rangle$ with $|-15/2\rangle$. If
the basic exchange process (i.e.\ the one for which $\Delta M_S=0,\,\pm 1$)
occurs in, say, $k$th order\footnote{For example, ``potential exchange'' in 1st
order, ``kinetic exchange'' in 2nd order, etc.} of perturbation theory then an
off-diagonal matrix element between the two components can only appear in
$(5k)$th order of perturbation theory. It is therefore reasonable to assume
that the off-diagonal matrix element is negligibly small compared to the
diagonal matrix elements ($\Delta M_S=0$) so that the effect of exchange
interaction on the Kramers doublet is accurately described by the $s_z$ spin
operator only.

Note that we derived the selection rule for exchange interaction on the basis
of the spin quantum number $M_S$ only, without paying attention to the angular
momentum quantum number $M_L$, although $M_L$ changes even more than $M_S$
between the states of the Kramers doublet \eqref{4:eq:KD}. The existence and
precise form of a selection rule for $M_L$ depends on the spatial symmetry of
the exchange problem under consideration. $\Delta M_L$ is therefore not as
useful as $\Delta M_S$ for predicting the vanishing of certain matrix elements
of exchange interaction. Note however that, even in the lowest symmetry, there
is a maximum to the amount that $M_L$ can change in the basic exchange process
described in the previous paragraph: within the $f$ orbitals, a one-electron
process can bring about at most a change of $\Delta M_L=\pm 6$. So we would
have, in general, that a basic exchange process can induce the following
changes in a lanthanide state: 
\begin{equation}\label{4:eq:selectionrule}
\begin{split}
&\Delta M_S=0,\,\pm 1\\
&|\Delta M_L|\leq 6.
\end{split}
\end{equation}
Thus at least two steps of this kind are needed to bridge $\Delta M_L=\pm 10$, 
but at least five are needed to bridge $\Delta M_S=\pm 5$. So in this case, the
selection rule of $M_S$ gives the stronger result and leads to the
conclusions---reached in the last paragraph---on the matrix elements of
exchange interaction in the Kramers doublet \eqref{4:eq:KD}.

We can now derive the precise form of that part of the effective Hamiltonian
that refers to the exchange interaction between a \ce{Dy^3+} ion [with ground
state \eqref{4:eq:KD}] and another magnetic center. We consider two cases,
which we shall encounter in the examples in Sections \ref{4:sec:chain} and
\ref{4:sec:ring}: in the first case the other center is another \ce{Dy^3+} ion;
in the second case the other center is an ion with an isotropic spin moment
$\mathbf{S}$. Consider first the exchange interaction with another \ce{Dy^3+}
ion. We assume that the second \ce{Dy^3+} has the same property of uniaxiality
as the first one and that it also shares all other relevant properties
discussed in the previous paragraphs. Then both ions are represented by a
spin-1/2 doublet with a local anisotropy axis $\hat{\mathbf{z}}_i$ ($i=1\,,2$)
and we already know that the effect of exchange interaction in each doublet is
proportional to $s_{z_i}$. Therefore the exchange Hamiltonian is necessarily 
of the form 
\begin{equation}\label{4:eq:exch_DyDy}
-J s_{z_1} s_{z_2},
\end{equation}
where it is understood that the first spin belongs to ion 1 and the second to
ion 2. Note that $\hat{\mathbf{z}}_1$ and $\hat{\mathbf{z}}_2$ need not be
parallel with each other. 

As a second case, consider the interaction between
\ce{Dy^3+} (anisotropy axis $\hat{\mathbf{z}}_1$) and an isotropic spin
$\mathbf{S}_2$. The latter may typically be a transition metal ion with
quenched orbital momentum. We found that the exchange processes that contribute
do not change the spin projection on the \ce{Dy^3+} ion ($\Delta M_{S_1}=0$,
quantization axis $\hat{\mathbf{z}}_1$), and that this result is independent of
the exchange partner. It is also known that every exchange process commutes
with the total spin (the matrix elements involved are spin-independent matrix
elements of kinetic, potential, and Coulomb energy \cite{PhYsRev_115_2}), so that
$\Delta M_{S_1}+\Delta M_{S_2}=0$. It follows then that $\Delta M_{S_2}=0$, or,
the exchange Hamiltonian commutes with the $z_1$ component of $\mathbf{S}_2$.
The simplest expression compatible with this requirement is
\begin{equation}\label{4:eq:exch_DyS}
-J s_{z_1}S_{z_1},
\end{equation}
where $s$ naturally represents \ce{Dy^3+} and $S$ represents $\mathbf{S}_2$.
The interaction is of Ising form with the anisotropy axis of \ce{Dy^3+} as
Ising axis. Note that, when $S_2>1/2$, higher powers of $S_{z_1}$ may enter 
the Hamiltonian. Considering exchange interaction as a perturbation however,
one can usually assume that the lowest-order contribution,
Eq.~\eqref{4:eq:exch_DyS}, is the leading term. 

There are other cases conceivable, for example dipole-dipole interaction
between the moment of \ce{Dy^3+} and a neighboring moment. One will always
find, as above, that the Hamiltonian is a product of $s_z$ (belonging to
\ce{Dy^3+}) and a part that belongs to the other ion and whose form depends on
the kind of the other ion and on the details of the interaction.

We have now obtained that a \ce{Dy^3+} ion, if its ground state is
$|M_J=\pm15/2\rangle$, fairly well separated from excited states, interacts
with the magnetic field and with neighboring ions as an Ising spin-1/2, in the
sense that the interaction is always proportional to $s_z$, as expressed in the
Eqs.~\eqref{4:eq:zeeman}, \eqref{4:eq:exch_DyDy}, and \eqref{4:eq:exch_DyS}.
This means that a chain-like molecular structure having \ce{Dy^3+} ions of
this kind at regular positions in the chain would meet the requirements of a
decorated Ising chain, given in part by Eqs.~\eqref{4:eq:HamdecIsing} and
\eqref{4:eq:commutation}. It remains of course to be shown that
$|M_J=\pm15/2\rangle$ can indeed be the ground state of a coordinated
\ce{Dy^3+} ion in a polynuclear complex.

At first sight, this seems rather unlikely. $|M_J=\pm15/2\rangle$ is an
eigenstate of cylindrical symmetry. Within lanthanide $f^n$ states, the crystal
field is effectively cylindrical if there is, at least, an eightfold rotation
axis ($C_8$) or rotation-inversion axis ($S_8$).  \cite{AbRagam_EPR} $S_8$
symmetry has been obtained, for example, in mononuclear bis(phtalocyaninato)
sandwich complexes of the lanthanides.  \cite{IC_42_2440} Even when such high
symmetry is attained, the ground state is not necessarily the cylindrical
doublet with highest $|M_J|$ value.  \cite{IC_42_2440} Apart from that, the
symmetry of the coordination sphere of a lanthanide ion in a polynuclear,
possibly heterometallic, complex or chain is usually much lower or even
completely absent. Such is the case for the two examples considered in this
paper. On the basis of symmetry alone, there is thus no reason to expect
$|M_J=\pm15/2\rangle$ to be an eigenstate, let alone the ground state.
Nevertheless, recent \textit{ab initio} calculations have revealed the
unexpected result that the ground state of several low-symmetry complexes of
\ce{Dy^3+} is very close to $|M_J=\pm15/2\rangle$.
\cite{AngewChemIntEd_47_4126,NJChem_33_1224,ChemEurJ_15_11808,
AngewChemIntEd_49_7583,JACS_131_5573} They used the multiconfigurational,
wavefunction-based CASSCF/RASSI-SO method, to obtain accurate wavefunctions for
several of the lowest Kramers doublets.  Calculation of the principal $g$
factors of these states gives an indication of their composition. It was found
in several cases that the ground doublet has $g_z$ close to, but lower than 20,
and $g_x$ and $g_y$ close to 0.  This corresponds to a doublet mainly composed
of $|M_J=\pm15/2\rangle$. 

With evidence of \textit{ab initio} calculations it is thus possible to
identify \ce{Dy^3+} in certain coordination environments as an Ising spin-1/2
(to a good approximation). We will use this information to identify the
compounds in Sections \ref{4:sec:chain} and \ref{4:sec:ring} as decorated Ising
chains (or rings).

To conclude this section we remark that one cannot deduce from the vanishing
of two $g$ factors alone that a Kramers doublet will behave as an Ising spin.
It will, of course, in its interaction with the magnetic field
[Eq.~\eqref{4:eq:zeeman}], but this will not, in general, be true for the
exchange interaction. Take again the example of \ce{Dy^3+}, supposing the
ground state is the Kramers doublet $|M_J=\pm 7/2\rangle$. It is perfectly
uniaxial because $g_x=g_y=0$ and $g_z=9\frac{1}{3}$. A closer look at the
expansion of $|M_J=\pm 7/2\rangle$ in terms of the Russel-Saunders states
$|M_L\rangle|M_S\rangle$ shows, however, that the selection rules in
Eq.~\eqref{4:eq:selectionrule} permit a matrix element to exist between 
$|+7/2\rangle$ and $|-7/2\rangle$, therefore introducing non-Ising terms
(i.e., $s_x$ and $s_y$) in the same order of perturbation theory as the Ising
term in the exchange Hamiltonian.

\subsection{Nature of the eigenstates and level
crossings}\label{4:sec:eigenstates}

There exists a similarity between the spectra of the decorated and undecorated
Ising chains at which we want to take a closer look here. We suppose infinite,
periodic chains, or periodic, even-membered rings (In odd-membered rings
spin-frustration complicates the picture). We also limit ourselves to chains
with Ising spins of 1/2. 

The undecorated, or simple, Ising chain in a magnetic field $B$ parallel with
the $z$ axis is given by Eq.~\eqref{4:eq:HamdecIsing} and
\begin{equation}\label{4:eq:Ham_Ising}
\hat{h}_i=-Js_is_{i+1}-\mu_\mathrm{B}gs_iB, \end{equation} where $s_i$ is the
$z$ component of $\mathbf{s}_i$ ($s_i=\pm 1/2$). We assume, without loss of
generality, $B\geq0$.  The eigenstates of the chain are spin
\textit{configurations} like
($\uparrow\uparrow\downarrow\uparrow\downarrow\,$\ldots), etc. Of the $2^n$
eigenstates only two distinct ones can be the ground state: the ferromagnetic
(F) and the antiferromagnetic (AF): \begin{equation}\label{4:eq:Ising_GS}
\begin{split} \text{F}&:(\uparrow\uparrow\uparrow\uparrow\ldots)\quad
\text{if}\mspace{13mu} J>0 \quad \text{or} \quad J<0 \mspace{9mu} \text{and}
\mspace{9mu} \mu_\mathrm{B}gB>|J|,\\
\text{AF}&:(\uparrow\downarrow\uparrow\downarrow\ldots)\quad
\text{if}\mspace{13mu} J<0
\mspace{9mu}\text{and}\mspace{9mu}\mu_\mathrm{B}gB<|J|.\end{split}\end{equation} 
When $B=0$ time-reversal symmetry makes every state degenerate with the state
formed by flipping all the spins. This degeneracy is meant to be implied in
\eqref{4:eq:Ising_GS}, where only one of two states is shown in each case. Only
the two AF states remain degenerate when $B\neq0$.  When $J<0$ a ground state
level crossing occurs from AF to F when $B$ is increased. At the point of
crossing ($\mu_\mathrm{B}gB=|J|$) the two AF states are degenerate together
with all states derived from an AF state by flipping one or more down-spins up.
However, no other state than AF or F can be the ground state at any other value
of $B$. Thus the ground state of the Ising chain is either F or AF, and they
are degenerate, together with an infinite number of other states, at the
crossing point.

We now decorate the Ising bonds with identical but arbitrary units to obtain a
periodic decorated Ising chain. The spectrum is given by
Eq.~\eqref{4:eq:energies}. Since the chain is periodic, the spectrum of the
individual units $\hat{h}_i$ is independent of $i$ and the energies may be
written as $\varepsilon_k$, with $k$ ranging over the eigenstates of
$\hat{h}_i$. There are four sets of $\varepsilon_k$:
$\varepsilon_k(\uparrow\uparrow)$, $\varepsilon_k(\uparrow\downarrow)$,
$\varepsilon_k(\downarrow\uparrow)$, and $\varepsilon_k(\downarrow\downarrow)$,
in an obvious notation. Notice that the eigenstates of this chain can still be
classified according to the configuration of the \textit{Ising} spins:
($\uparrow\uparrow\downarrow\uparrow\downarrow\,$\ldots) etc., which follows
from the fact that all the $s_i$ and $\hat{H}$ form a commuting set of
observables. 

An interesting question is whether the same rules hold for the ground state of
the decorated Ising chain as did for the simple Ising chain. The answer is yes;
the ground state is either F or AF (referring to the Ising spin configuration)
and a crossing between them is possible, with the same number and kind of
degenerate states as in the simple Ising chain. To show this, we have to
consider only the lowest eigenstate belonging to each of the $2^n$ possible
Ising spin configurations. In these states every unit is in its lowest possible
state for the given orientation of the neighboring Ising spins:
$\varepsilon_1(s,s')$ (we assume this energy to be nondegenerate), so
that the total energy of the chain state is
\begin{equation}\label{4:eq:energy_chain}
E=n_{\uparrow\uparrow}\varepsilon_1(\uparrow\uparrow)+
n_{\downarrow\downarrow}\varepsilon_1(\downarrow\downarrow)+
n_{\downarrow\uparrow}\varepsilon_1(\downarrow\uparrow)+
n_{\uparrow\downarrow}\varepsilon_1(\uparrow\downarrow),
\end{equation}
where $n_{\uparrow\uparrow}$ denotes the number of pairs of neighboring Ising
spins that are both spin up, etc. For example, in the F configuration in 
Eq.~\eqref{4:eq:Ising_GS}, $n_{\uparrow\uparrow}=n$ (periodic boundary
conditions are assumed), while in the AF configuration,
$n_{\uparrow\uparrow}=0$. The eigenstates we have just described,  with energy 
\eqref{4:eq:energy_chain}, are in an obvious one-to-one correspondence with the
eigenstates of the simple Ising chain. The ground state is found by
minimizing \eqref{4:eq:energy_chain} with respect to the $n_{ss'}$, under the
restrictions
\begin{equation}\label{4:eq:restrictions}
\begin{split}
n_{\uparrow\uparrow}+n_{\downarrow\downarrow}+n_{\downarrow\uparrow}+
n_{\uparrow\downarrow}&=n\\
n_{\downarrow\uparrow}&=n_{\uparrow\downarrow}.
\end{split}
\end{equation}
The first relation states that the total number of Ising spins (or,
equivalently, unit cells) is $n$. The second relation follows from the fact
that, in a cyclic spin configuration, every $\downarrow\uparrow$  pair must
eventually be followed by a $\uparrow\downarrow$ pair, possibly after a number
of $\uparrow\uparrow$ pairs. Another restriction is that whenever both
$n_{\uparrow\uparrow}$ and $n_{\downarrow\downarrow}$ are not zero,
$n_{\downarrow\uparrow}$ [and by Eq.~\eqref{4:eq:restrictions} also
$n_{\uparrow\downarrow}$] must be at least one. Using \eqref{4:eq:restrictions}
we can rewrite Eq.~\eqref{4:eq:energy_chain} as 
\begin{equation}\label{4:eq:energy_chain_mod}
\begin{split}
E&= n_{\uparrow\uparrow} \Bigl(\varepsilon_1(\uparrow\uparrow)-
\frac{1}{2}[\varepsilon_1(\uparrow\downarrow)
+\varepsilon_1(\downarrow\uparrow)]\Bigr)\\
&+n_{\downarrow\downarrow}\Bigl(\varepsilon_1(\downarrow\downarrow)-
\frac{1}{2}[\varepsilon_1(\uparrow\downarrow)
+\varepsilon_1(\downarrow\uparrow)]\Bigr)\\
&+\frac{n}{2}
[\varepsilon_1(\uparrow\downarrow)+\varepsilon_1(\downarrow\uparrow)],
\end{split}
\end{equation}
where we see that only the average 
$[\varepsilon_1(\uparrow\downarrow)+\varepsilon_1(\downarrow\uparrow)]/2$ of
the ``antiparallel'' energies enters the equation. The last term is a constant
and can be discarded for the purpose of relative energy considerations.

We can now derive the values of $n_{\uparrow\uparrow}$ and
$n_{\downarrow\downarrow}$ for the ground state of the chain, keeping in mind
that $\varepsilon_1(s,s')$ is a function of the magnetic field $\mathbf{B}$.
Suppose then, first, that $\mathbf{B}=0$. Time reversal symmetry asserts that
$\varepsilon_1(\uparrow\uparrow)=\varepsilon_1(\downarrow\downarrow)$ and
$\varepsilon_1(\uparrow\downarrow)=\varepsilon_1(\downarrow\uparrow)$. It is
simple to see that, depending on the relative ordering of
$\varepsilon_1(\uparrow\uparrow)$ and $\varepsilon_1(\uparrow\downarrow)$, $E$
is minimal in the F configuration ($n_{\uparrow\uparrow}=n$ or
$n_{\downarrow\downarrow}=n$; $n_{\uparrow\downarrow}=n_{\downarrow\uparrow}=0$
) when $\varepsilon_1(\uparrow\uparrow)<\varepsilon_1(\uparrow\downarrow)$ or
in the AF configuration ($n_{\uparrow\uparrow}=n_{\downarrow\downarrow}=0$;
$n_{\uparrow\downarrow}=n_{\downarrow\uparrow}=n/2$) when
$\varepsilon_1(\uparrow\uparrow)>\varepsilon_1(\uparrow\downarrow)$ (we exclude
the possibility of equality of both energies from the discussion;
$\varepsilon_1(\uparrow\uparrow)=\varepsilon_1(\uparrow\downarrow)$ would
correspond, in the simple Ising chain \eqref{4:eq:Ham_Ising}, with $J=0$).
When $\mathbf{B}\neq 0$, time reversal symmetry is not operative, and we have,
in general, four different energies $\varepsilon_1(s,s')$. The equation
\eqref{4:eq:energy_chain_mod} shows that the configuration that minimizes $E$
is determined by the sign of the two terms in round brackets; if both are
positive, then $n_{\uparrow\uparrow}=n_{\downarrow\downarrow}=0$ (AF
configuration); if at least one of them is negative, then either
$n_{\uparrow\uparrow}=n$ or $n_{\downarrow\downarrow}=n$ (F configuration),
depending on whether respectively $\varepsilon_1(\uparrow\uparrow)$ or
$\varepsilon_1(\downarrow\downarrow)$ is lower. 

Finally, the magnetic field can induce a transition from the AF to an F ground
state configuration, say with all Ising spins up. This happens when
\begin{equation}\label{4:eq:AF_F_transition} \varepsilon_1(\uparrow\uparrow)=
\frac{1}{2}[\varepsilon_1(\uparrow\downarrow)
+\varepsilon_1(\downarrow\uparrow)], \end{equation} and
$\varepsilon_1(\uparrow\uparrow)<\varepsilon_1(\downarrow\downarrow)$. At this
point, the ground state configurations are all those for which
$\{n_{\downarrow\downarrow}=0, n_{\uparrow\uparrow}= 0,2,4,\ldots,n\}$, exactly
the same as in the simple Ising chain.

We find thus a complete analogy between the simple and the decorated Ising
chain as far as the ground state Ising spin configuration is concerned. The
only possible configurations are the fully antiferromagnetically aligned and
the fully ferromagnetically aligned configurations. No ``intermediate''
configuration can be the ground state. The only exception is the crossing point
between AF and F, where there is a high degeneracy of configurations. These
conclusions are independent of the nature of the decorating unit.

Although the decorated Ising model predicts that the AF and F ground states are
both doubly degenerate (in zero field), this degeneracy is not a result of the
spatial symmetry: in the cyclic group $C_n$, the two AF components combine into
irreducible representations (irreps) $A$ and $B$, while the two F components
transform as two $A$ irreps. Introduction of neglected terms in the
Hamiltonian, that destroy the Ising property, could split these ground state
components.

The decorated Ising chain affords two new kinds of ground state level
crossings, not present in the simple Ising chain. The first of these is the
transition between one F configuration of the Ising spins and the other:
$(\uparrow\uparrow\uparrow\uparrow,\ldots)\leftrightarrow
(\downarrow\downarrow\downarrow\downarrow,\ldots)$. This transition takes place
when $\varepsilon_1(\uparrow\uparrow)=\varepsilon_1(\downarrow\downarrow)$.
This level crossing, induced by the magnetic field, can for example be
encountered in frustrated Ising-Heisenberg
chains.\cite{JPhysCondensMatter_18_4967} A second kind of new ground state
transition arises from the crossing of levels \textit{within} a decorating
unit. A ground state crossing can result in which the Ising spin configuration
remains the same but the state corresponding to $\varepsilon_1(s,s')$ crosses
with the state corresponding to $\varepsilon_2(s,s')$. More precisely this
happens in the F configuration when
$\varepsilon_1(\uparrow\uparrow)=\varepsilon_2(\uparrow\uparrow)$ and in the AF
configuration when either
$\varepsilon_1(\uparrow\downarrow)=\varepsilon_2(\uparrow\downarrow)$ or
$\varepsilon_1(\downarrow\uparrow)=\varepsilon_2(\downarrow\uparrow)$, or both.
The previous paragraphs have shown that we do not need to consider 
configurations other than F and AF for the ground state. 

Level crossings are usually connected with the presence of good quantum
numbers. For the Ising-type crossings, the relevant conserved quantities are
the $n$ Ising spins $\{s_i\}$. The crossing of energy levels within the
decorating unit should be associated with a conserved variable that is
\textit{internal} to that unit, much the same as in isolated molecules.
In Section \ref{4:sec:chain} we will
encounter an example where both transitions---Ising type and internal
type---occur in a magnetic field.

\subsection{Magnetization of powder samples}

In the following sections we will be comparing our theoretical results with
measurements performed on powder samples of the crystalline compounds. In this
section we consider the powder averaging of magnetization for the example of
the simple Ising chain.

Let $\theta$ and $\phi$ be the polar
angles of the magnetic field vector with respect to a molecular reference
frame, then the free energy is a function of $\theta$, $\phi$, and the strength
of the field, $B$: $f(\theta,\phi,B)$. The projection of the magnetization on
the field direction $\hat{\mathbf{e}}_B$ is
\begin{equation}\label{4:eq:magnprojection}
\hat{\mathbf{e}}_B\cdot\mathbf{M}(\theta,\phi,B)=-\frac{\partial
f(\theta,\phi,B)}{\partial B}.
\end{equation}
Averaging over one hemisphere gives the powder
magnetization
\begin{equation}\label{4:eq:powdermagnetization}
M(B)=\frac{1}{2\pi}\int_0^{2\pi}\int_0^{\pi/2}-\frac{\partial
f(\theta,\phi,B)}{\partial B}\sin\theta\, d\theta\, d\phi.
\end{equation}

Let us see how the powder averaging affects the magnetization curve for a
simple Ising chain. Take a spin-1/2 infinite antiferromagnetic Ising chain
with anisotropy axes parallel with each other and with the $z$ axis, and
uniaxial $g$-factors ($g_x=g_y=0$ and $g_z\equiv g$). This could for example be
realized by a chain of identical \ce{Dy^3+} units (see Section \ref{4:sec:Dy}).
The Hamiltonian is given by Eq.~\eqref{4:eq:HamdecIsing}, substituting
\begin{equation}\label{4:eq:Ham_Ising_2}
\hat{h}_i=-Js_is_{i+1}-\mu_\mathrm{B}gs_iB\cos\theta,
\end{equation}
where $B\cos\theta$ is the $z$ component of the magnetic field, and $J<0$.
Defining $j\equiv J/k$ and $b \equiv \mu_\mathrm{B}gB/k$, the magnetization,
which has only a nonzero $z$ component, is\cite{Yeomans} 
\begin{equation*}
M_z=\frac{\mu_\mathrm{B}g}{2}\frac{\sinh[
b\cos\theta/2T]}{\sqrt{\sinh^2[b\cos\theta/2T]+e^{-j/T}}}.
\end{equation*}
The projection on the field direction [Eq.~\eqref{4:eq:magnprojection}] is
$M_z\cos\theta$. Plugging this in Eq.~\eqref{4:eq:powdermagnetization} and
substituting $u=\cos\theta$ yields the powder magnetization of the Ising chain
\begin{equation}\label{4:eq:powdermagIsing}
M=\frac{\mu_\mathrm{B}g}{2}\int_0^1\frac{\sinh[
b u/2T]}{\sqrt{\sinh^2[b u/2T]+e^{-j/T}}}u\,du.
\end{equation}
Because the Hamiltonian in Eq.~\eqref{4:eq:Ham_Ising_2} does not depend on
$\phi$, this variable has been integrated out in
Eq.~\eqref{4:eq:powdermagIsing}. 

\begin{figure}
\centering
\includegraphics{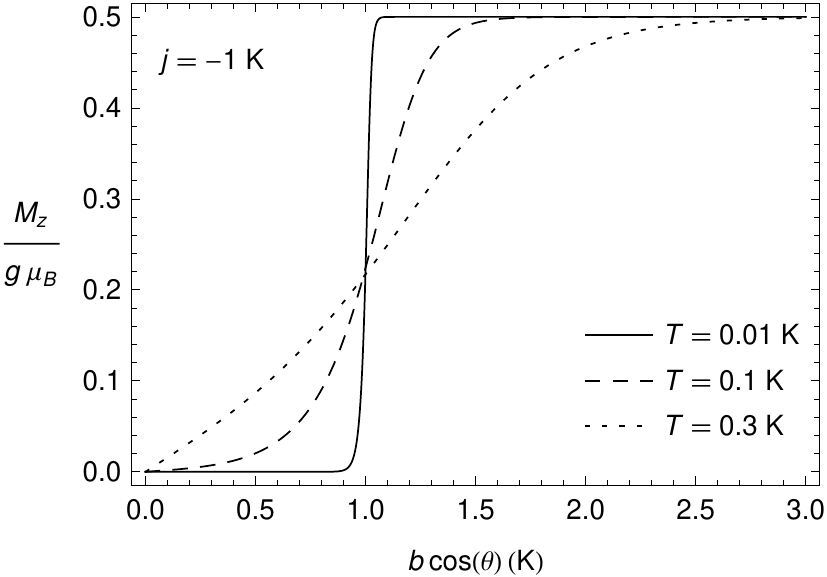}
\caption{Magnetization along the $z$ direction versus field of
the antiferromagnetic Ising chain, defined in Eq.~\eqref{4:eq:Ham_Ising_2}. 
The curve approaches a perfect step as
$T\rightarrow 0\,\mathrm{K}$. \label{4:fig:magnIsing_zas}}
\vspace{2cm}
\includegraphics{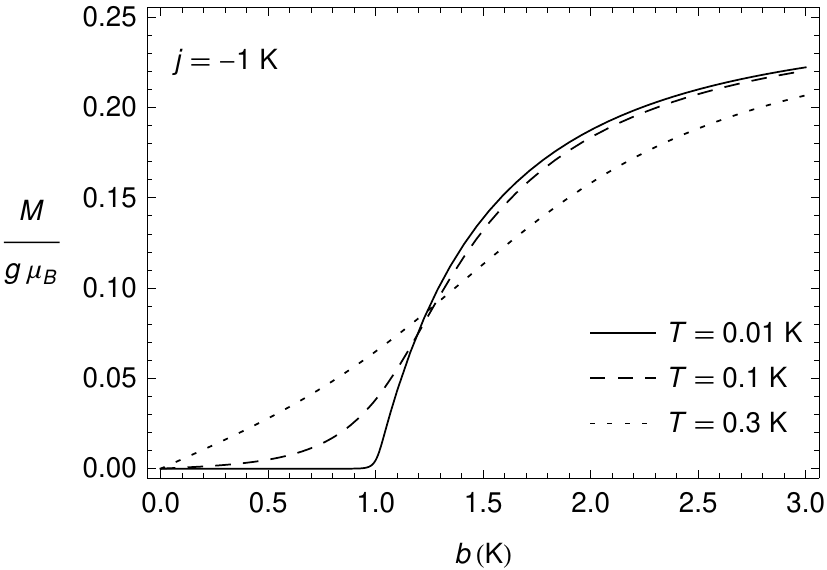}
\caption{Magnetization of a powder sample of antiferromagnetic Ising chains. 
The low-temperature limiting curve is given in 
Eq.~\eqref{4:eq:powdermagIsing0K}.
\label{4:fig:magnIsing_powder} }
\end{figure}

Figs.~\ref{4:fig:magnIsing_zas} and \ref{4:fig:magnIsing_powder} show plots of
magnetization versus magnetic field, for coupling constant
$j=-1\,\mathrm{K}$. The step-like appearance of $M_z$ is associated with the
ground state crossing that occurs at $b\cos\theta=\abs{j}$. At that point, the
antiferromagnetic ground state (or rather ground state Ising doublet) is
replaced by the ferromagnetic state (all spins up). Consequently, the
magnetization jumps from zero to the saturation value of 0.5, as seen in
Fig.~\ref{4:fig:magnIsing_zas}. The magnetization of a powder sample of the 
same Ising chain is shown in Fig.~\ref{4:fig:magnIsing_powder}. Before the
crossing point, $M$ behaves qualitatively the same as $M_z$. After the
crossing point however, $M$ is seen to reach only slowly its saturation
value, which is half of the saturation value of $M_z$, viz.\ 0.25. The limiting
curve of $M$ as $T\rightarrow 0\,\mathrm{K}$ can be calculated exactly from
Eq.~\eqref{4:eq:powdermagIsing}:
\begin{equation}\label{4:eq:powdermagIsing0K}
M\Big\vert_{T\rightarrow 0}=\begin{cases}
0& \text{if $b\leq\abs{j}$}\\
{\displaystyle\frac{\mu_\mathrm{B} g}{4}\left(1-\frac{\abs{j}^2}
{b^2}\right)}& \text{if $b>\abs{j}$}.
\end{cases}
\end{equation}
(This is of course only valid for the antiferromagnetic case $j<0$.)
In Fig.~\ref{4:fig:magnIsing_powder}, this limiting curve is very closely 
approximated by the curve at $T=0.01\,\mathrm{K}$. Clearly, the sharp step of
$M_z$ transforms in the powder to the concave form displayed by $M$. This is
understood from the fact that, in a powder, for a given field $b>\abs{j}$,
there is always a fraction of molecules that is not magnetized (in the sense
that they are in the antiferromagnetic ground state) because they are oriented
so with respect to the field, that $b\cos\theta<\abs{j}$ (see
Fig.~\ref{4:fig:magnIsing_zas}). The powder saturates only when every molecule
is fully magnetized, and this happens only for $b\rightarrow\infty$. Therefore,
$M$ (at $0\,\mathrm{K}$) does not abruptly saturate at the crossover point,
but increases slowly to saturation.

\subsection{Corrections for the contribution of excited Kramers doublets}
\label{4:sec:corrections}

In deriving the exchange Hamiltonian in Section \ref{4:sec:Dy} we
assumed that only the lowest Kramers doublet on \ce{Dy^3+} took part. This is
certainly a good approximation when the gap between the lowest and the
second-lowest Kramers doublet is much larger then the strength of the exchange
interaction. However, the excited Kramers doublets often have to be taken into
account to a certain degree of approximation if a comparison with experimental
data on susceptibility and magnetization is desired. The crystal field
splitting of the \ce{^6H_{15/2}} level is of the order of $kT$ at room
temperature. This gives rise to two effects: (i) a thermal population of
excited Kramers doublets, and (ii) a modification of the lowest Kramers doublet
as a function of the applied magnetic field by interaction with the excited
doublets. 

Effect (i) is mainly visible in the temperature dependence of $\chi T$, where
$\chi$ is the powder magnetic susceptibility: for a single \ce{Dy^3+} center,
$\chi T$ increases monotonically with increasing temperature, from the value of
the ground doublet at 0~K to the saturation value of \ce{^6H_{15/2}} at higher
temperatures. Effect (ii) gives rise to temperature-independent paramagnetism
(TIP). It is visible at temperatures sufficiently low so that only the ground
doublet is occupied. It contributes a linear increase of $\chi T$ with $T$ and
a linear increase of the magnetization $M$ with the applied field $B$. 

In the simplest approximation, the contribution of the excited Kramers doublets
to the magnetic properties of the chain is equal to the contribution they have
to the properties of the single, isolated \ce{Dy^3+} ion in the same ligand
environment it has in the chain. Let $\chi_\mathrm{DIC}$ and $M_\mathrm{DIC}$
denote susceptibility and magnetization derived from the decorated Ising chain
model, and let $\chi_\mathrm{Dy}$ denote the susceptibility of the \ce{Dy^3+}
center and $\mu'_\mathrm{Dy}B$ the magnetic moment induced by 
$B$ in the ground doublet of the \ce{Dy^3+} center, then the corrected
properties are (supposing one \ce{Dy^3+} ion per unit cell)
\begin{subequations}\label{4:eq:corrections}
\begin{align}
\chi T&= \chi_\mathrm{DIC}T +\Bigl(\chi_\mathrm{Dy} T
-\chi_\mathrm{Dy} T\big\vert_{T=0}\Bigr), \label{4:eq:corrections_chiT}\\
M &= M_\mathrm{DIC} + \mu'_\mathrm{Dy} B.
\label{4:eq:corrections_M}
\end{align}
\end{subequations}
The last equation assumes that only the ground doublet of \ce{Dy^3+} is
occupied. This is correct at the temperature at which magnetization curves are
usually recorded (e.g., 2\unit{K}).
If necessary, corrections due to other magnetic ions can be added in the same
way. $\chi_\mathrm{Dy}$ and $\mu'_\mathrm{Dy}$ can be obtained from
multiconfigurational \textit{ab initio} calculations
\cite{AngewChemIntEd_47_4126, NJChem_33_1224, ChemEurJ_15_11808,
AngewChemIntEd_49_7583}, or, in oligonuclear complexes, experimentally by
replacing certain magnetic ions with diamagnetic ions.\cite{ChemEurJ_4_1616}

The equations \eqref{4:eq:corrections} are evidently correct in the limit of
vanishing exchange interactions. The assumption we make is that the exchange
interactions are sufficiently small for these corrections to remain valid. A
more accurate approach should take into consideration the fact that the excited
Kramers doublets also participate in the exchange interaction with neighbors.
This is however out of the scope of the decorated Ising model.

\section{[D\lowercase{y}C\lowercase{u}M\lowercase{o}C\lowercase{u}]$_\infty$ 
polymer as Ising-Heisenberg chain}\label{4:sec:chain}

We now turn our attention to two actual examples of decorated Ising models
based on \ce{Dy^3+}: a chain, treated in this section, and a four-ring treated
in the following section. The problem is approached as follows: the Hamiltonian
for the chain in a magnetic field is formulated, with the help of the
considerations in Section \ref{4:sec:Dy}. Values of the $g$ factors of the
magnetic ions are taken directly from \textit{ab initio} calculations, reported
elsewhere.\cite{ChemEurJ_15_11808, AngewChemIntEd_49_7583} This leaves the
exchange coupling constants as parameters of the model, to be fitted by
comparison with experimental magnetization and susceptibility data. (In Section
\ref{4:sec:ring}, the direction of the anisotropy axis contributes one extra
parameter.)

\begin{figure}
\centering
\includegraphics{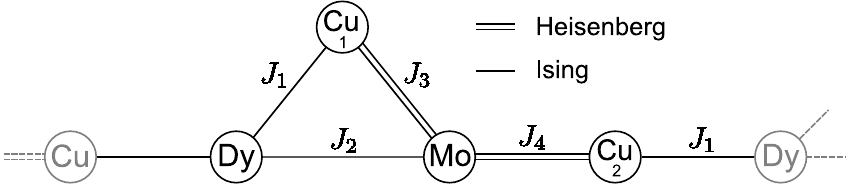}
\caption{Scheme of one unit cell (black) of the [DyCuMoCu]$_\infty$ chain,
showing type of exchange interactions and labeling of exchange
constants. See Ref.~\onlinecite{ChemEurJ_15_11808} for the complete molecular
structure.\label{4:fig:chainscheme}}
\end{figure}

The [DyCuMoCu] chain was recently synthesized and details of its chemical
composition and structure are given in Ref.~\onlinecite{ChemEurJ_15_11808}. The
crystal structure was found to consist of parallel linear chains each made of
[DyCuMoCu] unit cells.  Fig.~\ref{4:fig:chainscheme} shows how the metal ions
are connected by ligand bridges. Multiconfigurational CASSCF/RASSI-SO
calculations have been performed on each of the four metal ions in their ligand
environment, suitably disconnected from the rest of the chain (details of the
calculations can be found in Ref.~\onlinecite{ChemEurJ_15_11808} and the
accompanying Supplementary Information). Most important for us is that the
\ce{Dy^{3+}} center was found to have a ground Kramers doublet, separated by
141 cm$^{-1}$ from the second
doublet, and characterized by complete uniaxial anisotropy:
\begin{equation}\label{4:eq:Dy_gfactors_chain}
\mathrm{Dy^{3+}}: \qquad g_z=19.6,\quad g_x=0, \quad g_y=0.
\end{equation}
(Actually $g_x$ and $g_y$ were calculated about 0.03, which is small enough to
be ignored.) The value of $g_z=19.6$ shows that this doublet is only slightly
perturbed from the $|M_J=\pm15/2\rangle$ doublet of the \ce{^6H_{15/2}} level,
the latter having $g_z=20$. Together with the fact that the energy gap to the
second Kramers doublet is about ten times larger than the exchange interaction
(as we will find later), these results indicate that the \ce{Dy^{3+}} ion will
behave as an Ising spin, as described in Section \ref{4:sec:Dy}. As a side-note
we may add that the total splitting of the \ce{^6H_{15/2}} level was calculated
to be 560 cm$^{-1}$, which is indeed of the order of room-temperature $kT$ (see
Section \ref{4:sec:corrections}).

Both \ce{Cu^{2+}} ($d^9$) and \ce{Mo^{5+}} ($d^1$) have a spin = 1/2, orbitally
nondegenerate ground state, well separated ($> 15 000\, \mathrm{cm}^{-1}$) from
higher states. The two \ce{Cu^{2+}} ions in the unit cell reside in almost
identical environments\cite{ChemEurJ_15_11808} and have therefore virtually the
same properties. The calculated $g$ factors are tetragonal:
$g_{\|}=2.33,\,g_\bot=2.07$ for \ce{Cu^{2+}} and $g_\|=2.00,\,g_\bot=1.95$ for
\ce{Mo^{5+}}. To avoid unnecessary complications we will regard these ions as
isotropic spins with root-mean-square $g$ factors
\begin{equation*}
\begin{split}
\mathrm{Cu^{2+}}&:\qquad g_\mathrm{Cu}=2.16,\\
\mathrm{Mo^{5+}}&:\qquad g_\mathrm{Mo}=1.97.
\end{split}
\end{equation*}
This approximation will not have important consequences for the magnetic
properties, which are largely dominated by the high \ce{Dy^{3+}} moment anyway.

We introduce exchange interaction between metal ions directly connected by
ligand bridges. \ce{Dy^{3+}} interacts with its three neighbors via the Ising
Hamiltonian Eq.~\eqref{4:eq:exch_DyS}. Interaction between the isotropic spins
is given by the Heisenberg Hamiltonian $-J_{ij}\mathbf{S}_i\cdot\mathbf{S}_j$.
Fig.~\ref{4:fig:chainscheme} shows the exchange configuration, with single
bonds representing Ising interaction and double bonds representing Heisenberg
interaction. Note that we have approximated the Dy-Cu$_1$ and Dy-Cu$_2$
coupling strengths to be equal ($J_1$), following the approximate local
symmetry of the Dy-Cu pairs.\cite{ChemEurJ_15_11808}

It is now possible to see that the [DyCuMoCu]$_\infty$ polymeric chain is
indeed an experimental realization of an Ising-Heisenberg chain where the
[CuMoCu] trimeric Heisenberg units decorate the Dy-Dy bonds and the
\ce{Dy^{3+}} Ising spins separate the [CuMoCu] Heisenberg trimers from each
other. The total Hamiltonian in a magnetic field $\mathbf{B}$ is then given by
Eq.~\eqref{4:eq:HamdecIsing} and
\begin{equation}\label{4:eq:Ham_chain}
\begin{split}
\hat{h}_i(s^z_i,s^z_{i+1})&= -J_1 s^z_i S^z_{i\mathrm{Cu}_1} 
-J_2 s^z_i S^z_{i\mathrm{Mo}}-J_3\mathbf{S}_{i\mathrm{Cu}_1}\cdot
\mathbf{S}_{i\mathrm{Mo}}\\
&\quad -J_4\mathbf{S}_{i\mathrm{Mo}}\cdot
\mathbf{S}_{i\mathrm{Cu}_2}
-J_1  S^z_{i\mathrm{Cu}_2}s^z_{i+1}\\
&\quad-\mu_\mathrm{B}g_\mathrm{Dy}s^z_iB_z
 -\mu_\mathrm{B} (g_\mathrm{Cu}\mathbf{S}_{i\mathrm{Cu}_1}
+g_\mathrm{Mo}\mathbf{S}_{i\mathrm{Mo}}\\
&\mspace{173mu} +g_\mathrm{Cu}\mathbf{S}_{i\mathrm{Cu}_2})\cdot\mathbf{B},
\end{split}
\end{equation}
where $s^z_i$ is shorthand for $s^z_{i\mathrm{Dy}}$ and $g_\mathrm{Dy}$ is the
$g_z$ factor of \ce{Dy^{3+}} [Eq.~\eqref{4:eq:Dy_gfactors_chain}]. The $z$ axis
is the anisotropy axis of the \ce{Dy^{3+}} center. We have not specified its
direction with respect to the chain axis but this is not important here because
there are no other axes in the problem (the \ce{Dy^{3+}} anisotropy axes are
parallel by translational symmetry and we have assumed \ce{Cu^{2+}} and
\ce{Mo^{5+}} isotropic). All the spins in Eq.~\eqref{4:eq:Ham_chain} are spins
of 1/2. 

The Hamiltonian exhibits some symmetry. It is rotationally invariant around
$z$, if $\mathbf{B}$ is rotated simultaneously. We may therefore restrict
$\mathbf{B}$ to lie in a plane through $z$, say the $xz$ plane. This simplifies
calculation of the powder magnetization Eq.~\eqref{4:eq:powdermagnetization}:
one has to integrate only over $\theta$.  When $\mathbf{B}$ is directed along
the $z$ axis, the $z$ component of the total spin \emph{in} each decorating
unit is conserved:
\begin{equation}\label{4:eq:Szintern}
S^z_i=S^z_{i\mathrm{Cu_1}}+S^z_{i\mathrm{Mo}}+ S^z_{i\mathrm{Cu_2}},\qquad
[S^z_i,\hat{H}(\mathbf{B}\parallel \mathbf{\hat{z}})]=0. 
\end{equation}
We also note that in
this case the Zeeman Hamiltonian commutes \emph{almost} with $\hat{H}$. It
would commute exactly when $g_\mathrm{Cu}=g_\mathrm{Mo}$, for then the last
term in Eq.~\eqref{4:eq:Ham_chain} reduces to $-\mu_\mathrm{B}g_\mathrm{Cu} 
S^z_iB_z$. 

We let the length of the chain go to infinity: $n\to\infty$. To solve for the
thermodynamic properties we are only required to find the eigenvalues
$\varepsilon_k (s,s')$ of Eq.~\eqref{4:eq:Ham_chain} (See Section
\ref{4:sec:decIsing}), with $k=1\ldots8$, corresponding to the $2^3$ possible
states of the [CuMoCu] spin unit. This is done by 4 numerical $8\times8$ matrix
diagonalizations, one for each $(s,s')$ pair.

\begin{figure}
\centering
\includegraphics{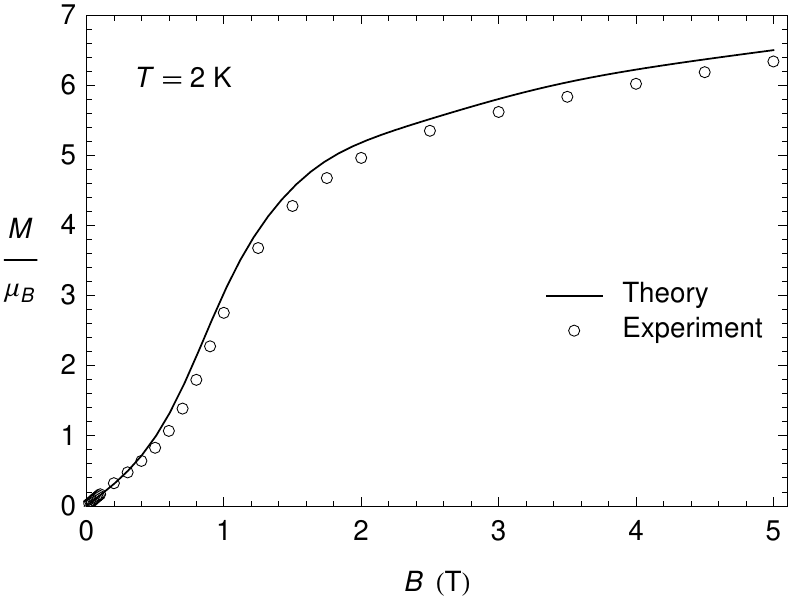}
\caption{Powder magnetization of [DyCuMoCu]$_\infty$.}
\label{4:fig:chainMp}
\end{figure}

We can now compare the theory with experiment. Powder magnetization (at
2\unit{K}) and susceptibility data have been recorded.\cite{ChemEurJ_15_11808}
We recall that we have to correct the theoretical curves before comparing with
experiment according to Eq.~\eqref{4:eq:corrections}. The corrections are
provided by the \textit{ab initio} calculations.\cite{ChemEurJ_15_11808}
$\mu'_\mathrm{Dy}$ turns out to be 0.03\unit{\mu_B/T}; the
accompanying correction in Eq.~\eqref{4:eq:corrections_M} is never more than
2.5\% of $M$. We ignore this correction. We do however correct $\chi$ as in
Eq.~\eqref{4:eq:corrections_chiT}.

\begin{figure}
\centering
\includegraphics{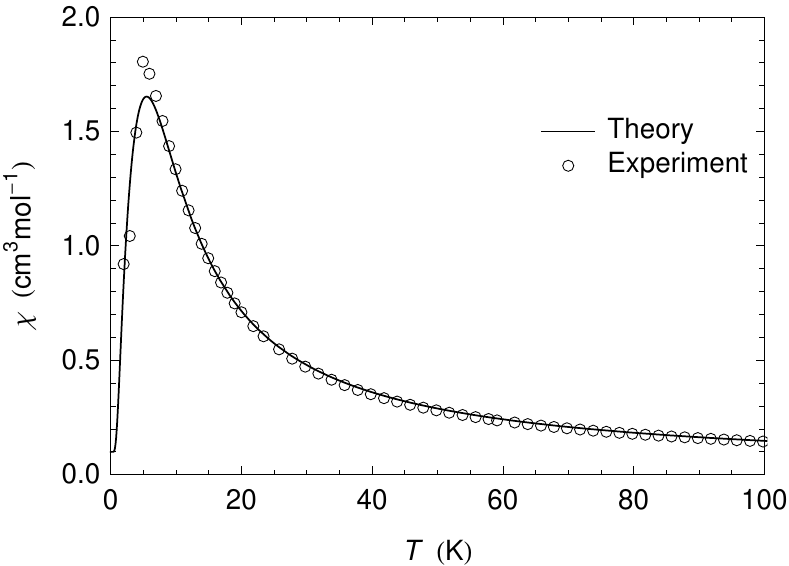}
\caption{Powder magnetic susceptibility of [DyCuMoCu]$_\infty$. The theoretical
curve contains the correction for the contribution of excited Kramers doublets.}
\label{4:fig:chainchip}
\end{figure}

Closest agreement with experiment was found for the following values of the
exchange constants (plots in Figs.~\ref{4:fig:chainMp} and
\ref{4:fig:chainchip}):
\begin{equation}\label{4:eq:J_chain}
\begin{split}
J_1&=15.3\unit{cm^{-1}},\; J_2=-8.0\unit{cm^{-1}}, \;
J_3=-8.3\unit{cm^{-1}},\\
J_4&= 11.8\unit{cm^{-1}}.
\end{split}
\end{equation}
These were obtained by a least-squares fit of $\chi$ followed by a small
manual adjustment to improve the fit of $M$  while not distorting that of
$\chi$ appreciably. (The least-squares fit of $\chi$ gave
$J_1=15.7\unit{cm^{-1}}$, $J_2=-8.3\unit{cm^{-1}}$, $J_3=-6.3\unit{cm^{-1}}$,
$J_4= 11.8\unit{cm^{-1}}$.) There are, as far as we know, no data in the
literature with which to compare the values in Eq.~\eqref{4:eq:J_chain}.
However, an experimental study is available of a (\ce{Dy^{3+}}, \ce{Cu^{2+}})
dinuclear complex in which the bridging ligand is the same as in this
chain.\cite{ChemEurJ_4_1616} The authors found a ferromagnetic interaction. A
superficial analysis of the susceptibility curve in that paper, using the Ising
Hamiltonian we use in this paper, yields $J_1=15\pm 5\unit{cm^{-1}}$. The other
values in \eqref{4:eq:J_chain} are difficult to assess. For a discussion of
these values and their relation with the molecular structure as well as some
evidence from DFT calculations, we refer to
Ref.~\onlinecite{ChemEurJ_15_11808}. Certainly, no confidence should be
attached to the numbers in decimal places in \eqref{4:eq:J_chain}.

\begin{figure}
\centering
\includegraphics{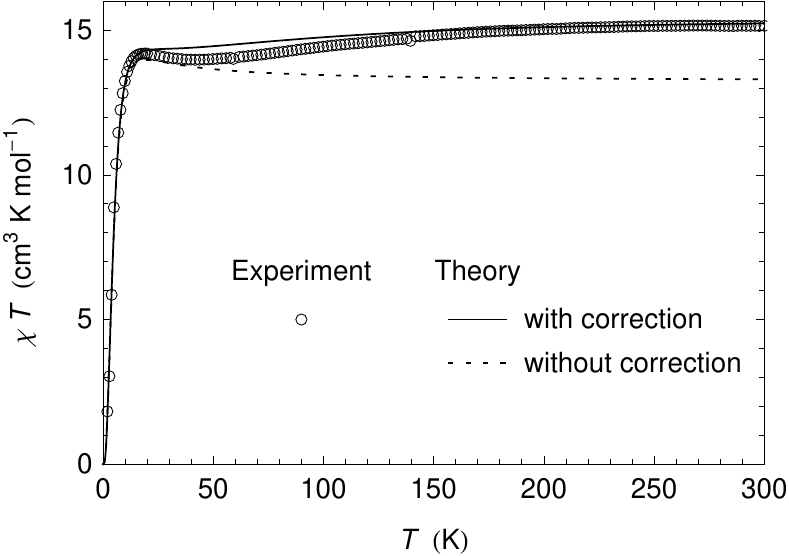}
\caption{Powder magnetic susceptibility of [DyCuMoCu]$_\infty$. The correction
refers to the contribution of excited Kramers doublets,
Eq.~\eqref{4:eq:corrections_chiT}.} 
\label{4:fig:chainchiTp}
\end{figure}

The effect of the excited Kramers doublets of \ce{Dy^{3+}} is most clearly seen
in the $\chi T$ curve (Fig.~\ref{4:fig:chainchiTp}). The curve shows a steady
increase above 50\unit{K} which is not predicted by our decorated Ising model,
but is indeed due to the thermal population of the Kramers doublets that
originate from the \ce{^6H_{15/2}} level. We can obtain the expected
high-temperature limit of $\chi T$ by considering the metal ions as independent
spins. The susceptibility components $\chi_{\alpha\alpha}$ of an angular
momentum multiplet $J$ with principal $g$-factors $g_\alpha$ ($\alpha=x,y,z$)
are given by \cite{kaHn_mol_magn}
\begin{equation}\label{4:eq:spinchi}
\chi_{\alpha\alpha}=\frac{N_A\mu_\mathrm{B}^2}{3kT}g_\alpha^2J(J+1).
\end{equation}
Summing over \ce{Dy^{3+}} (\ce{^6H_{15/2}}, $g_{15/2}=4/3$), \ce{Cu1}, \ce{Mo},
and \ce{Cu2} (all are isotropic) gives 
\begin{equation*}
\begin{split}
\chi T&=\frac{1}{3}\sum_\alpha\chi_{\alpha\alpha}T\\
      &=\frac{N_A\mu_\mathrm{B}^2}{3k}\left(g_{15/2}^2\frac{15}{2}\frac{17}{2}
+(g_\mathrm{Cu}^2+g_\mathrm{Mo}^2+
g_\mathrm{Cu}^2)\frac{1}{2}\frac{3}{2}\right)\\
      &=15.4\frac{\mathrm{cm^3\,K}}{\mathrm{mol}}.
\end{split}
\end{equation*}
A similar calculation, only including the lowest Kramers doublet of
\ce{Dy^{3+}}, with $g$-factors as in Eq.~\eqref{4:eq:Dy_gfactors_chain}, gives
13.2\unit{cm^3\,K\,mol^{-1}}.  The correction supplied by the \emph{ab initio}
calculations to account for this difference, is seen to cover nicely the
high-temperature part of the experimental curve. 

One notices that $\chi T$ shows a slight depression around 40\unit{K} which
is not entirely reproduced by the theory. This might indicate a failure of the
simple approximation we used to include the excited Kramers doublets.
Eq.~\eqref{4:eq:corrections_chiT} is certainly correct at very high
temperatures, when the exchange interactions are irrelevant, and at very low
temperatures, when the excited Kramers doublets are not occupied. If these two
regions do not overlap, however, there is a temperature window between, in
which excited doublets start to get occupied while exchange interaction is not
quite negligible yet. In that case, the exchange interaction of the occupied 
excited doublet(s) with other ions should be taken into account. Such an
interaction of antiferromagnetic type could possibly depress $\chi T$ as
observed. 

We shall now describe some features of the spectrum of the chain, paying
attention to the properties described in Section \ref{4:sec:eigenstates}.
Consider the chain without magnetic field. The exchange parameters in
Eq.~\eqref{4:eq:J_chain} predict a ground state that has an AF Ising spin
configuration. This is in accordance with the susceptibility measurements,
which show that $\chi T\to 0$ as $T\to 0$, requiring a nonmagnetic ground
state (Fig.~\ref{4:fig:chainchiTp}). The ground state is indeed nonmagnetic
because $|\varepsilon_1(\uparrow\downarrow)\rangle$ is the time-reversed state
of $|\varepsilon_1(\downarrow\uparrow)\rangle$. Let $M_S$ denote an
eigenvalue of $S^z$ [Eq.~\eqref{4:eq:Szintern}]: $M_S\in\{-3/2,-1/2,1/2,3/2\}$.
$S_i^z$ is conserved so $M_S$ may be used to label the eigenstates
$|\varepsilon_k(s,s')\rangle$ (we may leave out the index $i$ because all units
of the chain are identical). For the ground state, we find 
$M_S=-1/2$ in $|\varepsilon_1(\uparrow\downarrow)\rangle$ and $M_S=1/2$ in
$|\varepsilon_1(\downarrow\uparrow)\rangle$.

\begin{figure}
\centering
\includegraphics{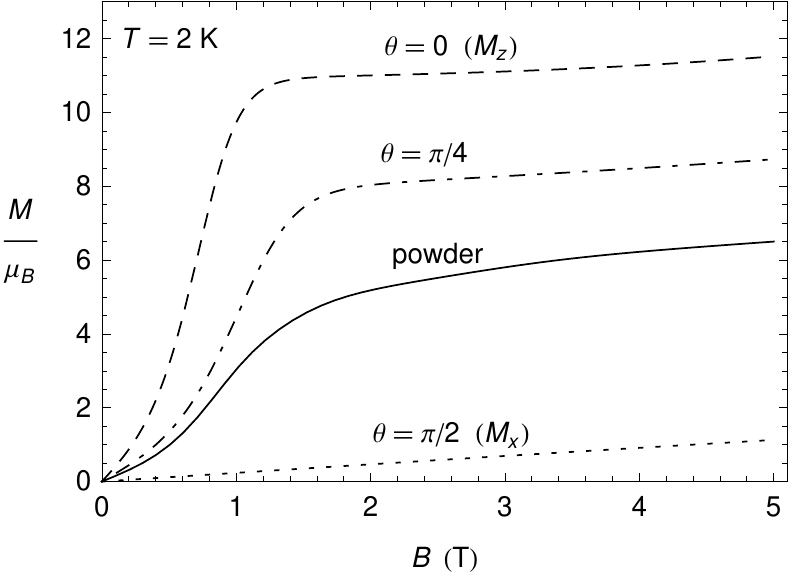}
\caption{Theoretical magnetization of [DyCuMoCu]$_\infty$. The powder
magnetization (See also Fig.~\ref{4:fig:chainMp}) is compared with
the projections of $\mathbf{M}$ on the field direction 
[Eq.~\eqref{4:eq:magnprojection}], for three different directions of the field;
$\theta$ is the angle between $\mathbf{B}$ and the $z$ axis.} 
\label{4:fig:chainMcomponents}
\end{figure}

\begin{figure}
\centering
\includegraphics{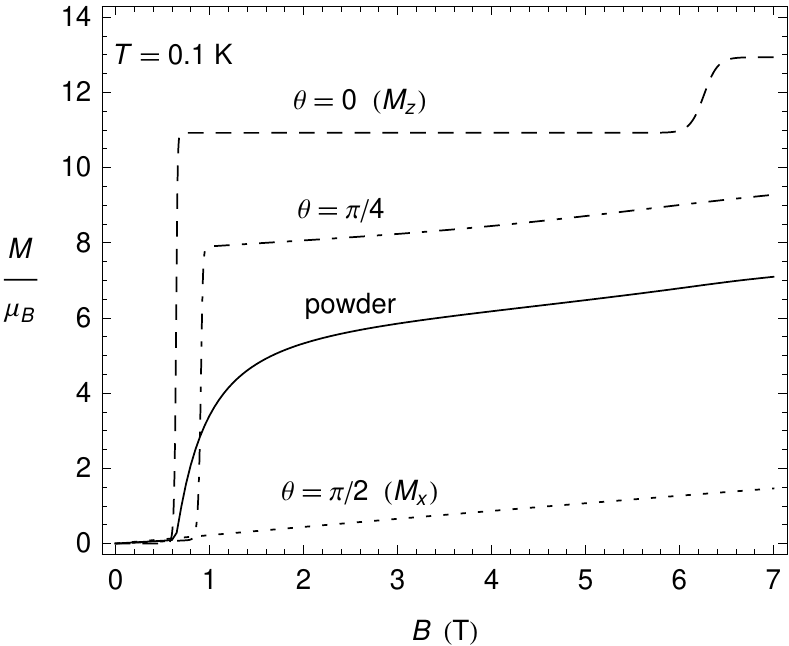}
\caption{Theoretical magnetization of [DyCuMoCu]$_\infty$. Same as
Fig.~\ref{4:fig:chainMcomponents} but at lower temperature and to higher
field.}
\label{4:fig:chainMcomponentsLowT}
\end{figure}

\begin{figure}
\centering
\includegraphics{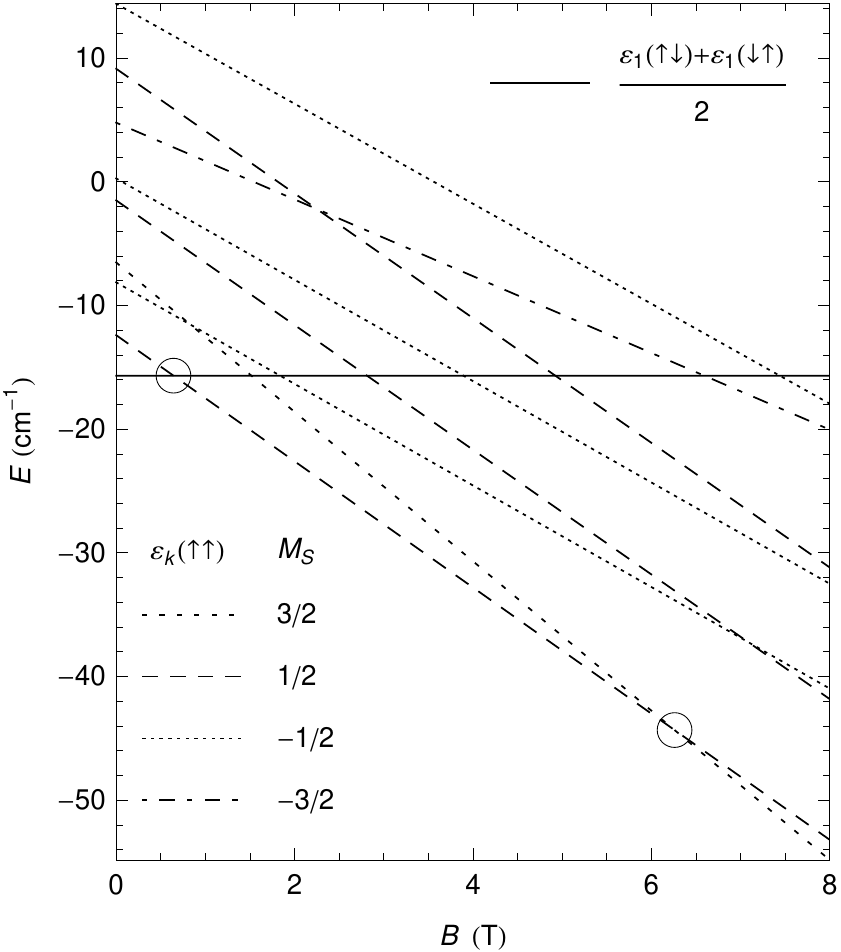}
\caption{Eigenvalues of $\hat{h}(s,s')$ [Eq.~\eqref{4:eq:Ham_chain}] in a
magnetic field \emph{parallel} with the $z$ axis ($\theta=0$). Circles indicate
ground state level crossings. The ground state of the chain is AF in zero field
(left), switches to F at 0.64\unit{T}, and undergoes an internal level crossing
at 6.3\unit{T}, marked by a change of the internal quantum number $M_S$ from
1/2 to 3/2. Both crossings can be seen in the $\theta=0$ magnetization curve in
Fig.~\ref{4:fig:chainMcomponentsLowT}. Note that the energy curves appear as
straight lines, although, with the exception of $M_S=\pm 3/2$, they are not
exactly straight, because the Zeeman Hamiltonian does not \emph{completely}
commute with the total Hamiltonian. All $\varepsilon_k(\uparrow\uparrow)$
decrease with increasing field strength because the large magnetic moment of
\ce{Dy^3+} dominates the smaller magnetic moments of the decorating unit.}
\label{4:fig:chainLevelcrossings} \end{figure}

Since the ground state is AF, we might expect that in a magnetic field a
crossover will occur to an F ground state. This is indeed what happens. The
convex increase of $M$ in Fig.~\ref{4:fig:chainMp} points to a flip of the
\ce{Dy^{3+}} spins to a parallel configuration. This is inferred from the value
of the magnetization, which approaches 6\unit{\mu_B} at 5\unit{T}. The [CuMoCu]
unit alone can only contribute a maximum of
$(2.16+1.97+2.16)/2=3.15\unit{\mu_B}$. The strong increase must come from the
contribution of the large \ce{Dy^{3+}} moments. 

The behavior of magnetization along certain directions of applied field is
shown in Fig.~\ref{4:fig:chainMcomponents}. The AF $\to$ F transition is most
clearly seen when the field is applied along $z$ ($\theta=0$); the transition
occurs below 1\unit{T}. After 1\unit{T}, $M_z$ reaches an approximately
constant plateau at $\approx 11\unit{\mu_B}$. The saturation
value of magnetization in direction $0\leq\theta\leq \pi/2$ is 
$(19.6\cos\theta+2.16+1.97+2.16)/2$. This gives $12.9\unit{\mu_B}$ for
$\theta=0$, which shows that $M_z$ has not quite reached its maximum at
5\unit{T} yet.

The positions of level crossings become more sharply defined on lowering the
temperature (Fig.~\ref{4:fig:chainMcomponentsLowT}). Here we also see that
$M_z$ undergoes a second transition at 6.3\unit{T}, after which it reaches
saturation. This transition is connected with a level crossing \emph{in} the
[CuMoCu] unit (see Section \ref{4:sec:eigenstates}) from $M_S=1/2$ to
$M_S=3/2$, as opposed to the first transition, at 0.64\unit{T}, which is of the
Ising type, described by Eq.~\eqref{4:eq:AF_F_transition}. The latter is the
analogue of the transition in the AF simple Ising chain
(Fig.~\ref{4:fig:magnIsing_zas}), while the ``internal'' transition has no such
analogue but is unique to the decorated Ising chain. The relevant energy level 
diagram is shown in Fig.~\ref{4:fig:chainLevelcrossings}. Note that, for fields
not parallel to $z$ (for example, $\theta=\pi/4$ in
Fig.~\ref{4:fig:chainMcomponentsLowT}), $M_S$ is not a quantum number and the
internal level crossing turns into an avoided crossing. This does not apply for
the Ising level crossing because the Ising spins are always conserved. Only
when the field is applied perpendicular to $z$ ($\theta=\pi/2$ in
Fig.~\ref{4:fig:chainMcomponentsLowT}) does the AF $\to$ F transition not occur
because the \ce{Dy^{3+}} spins do not interact with perpendicular fields.

\begin{figure}
\centering
\includegraphics{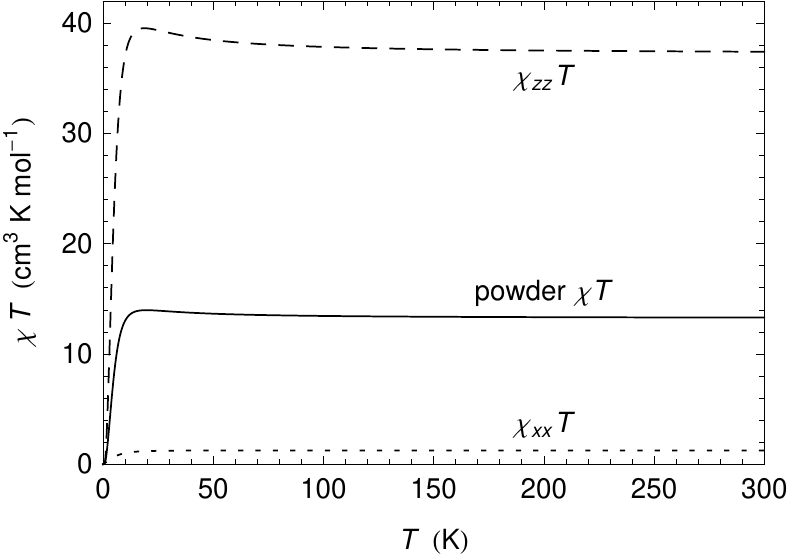}
\caption{Theoretical susceptibility of [DyCuMoCu]$_\infty$, without correction
for contribution of excited Kramers doublets. The powder $\chi T$ (see also
Fig.~\ref{4:fig:chainchiTp}) is compared with the Cartesian components of $\chi
T$. $z$ is the direction of the anisotropy axis of \ce{Dy^{3+}}, $x$ is any
direction perpendicular to $z$. $\chi=(\chi_{zz}+2\chi_{xx})/3$.}
\label{4:fig:chainchiTcomponents}
\end{figure}

The low-temperature limit of the powder magnetization in
Fig.~\ref{4:fig:chainMcomponentsLowT} may be compared with that of the simple
Ising chain in Fig.~\ref{4:fig:magnIsing_powder}. The resemblance is clear; the
decorated chain is different in the small linear increase of $M$ before the
transition, and the more linear approach to saturation, which lies at
($19.6/2+2.16+1.97+2.16)/2=8.0\unit{\mu_B}$. Both are due to TIP interaction in
the [CuMoCu] unit, the effect of which is most clearly seen in the $M_x$ curve
in Fig.~\ref{4:fig:chainMcomponentsLowT}.

To conclude this section we remark that the mentioned similarity with the
magnetization of the simple Ising chain is a consequence of the very high
magnetic moment of the \ce{Dy^{3+}} spins in comparison with the [CuMoCu] unit.
The dominance of \ce{Dy^{3+}} is most dramatically shown in the components of
$\chi T$ (Fig.~\ref{4:fig:chainchiTcomponents}). An application of
Eq.~\eqref{4:eq:spinchi} shows that the high-temperature limit of
$\chi_{xx}T$ is
$\frac{N_A\mu_\mathrm{B}^2}{3k}(2.16^2+1.97^2+2.16^2)\frac{1}{2}\frac{3}{2}=
1.24\unit{cm^3\,K\,mol^{-1}}$, while that of $\chi_{zz}T$ is
$\frac{N_A\mu_\mathrm{B}^2}{3k}(19.6^2+2.16^2+1.97^2+2.16^2)
\frac{1}{2}\frac{3}{2}=37.2\unit{cm^3\,K\,mol^{-1}}$.

\section{D\lowercase{y}$_4$C\lowercase{r}$_4$ complex as decorated Ising
ring}\label{4:sec:ring}

\begin{figure}
\centering
\includegraphics[width=8.6cm]{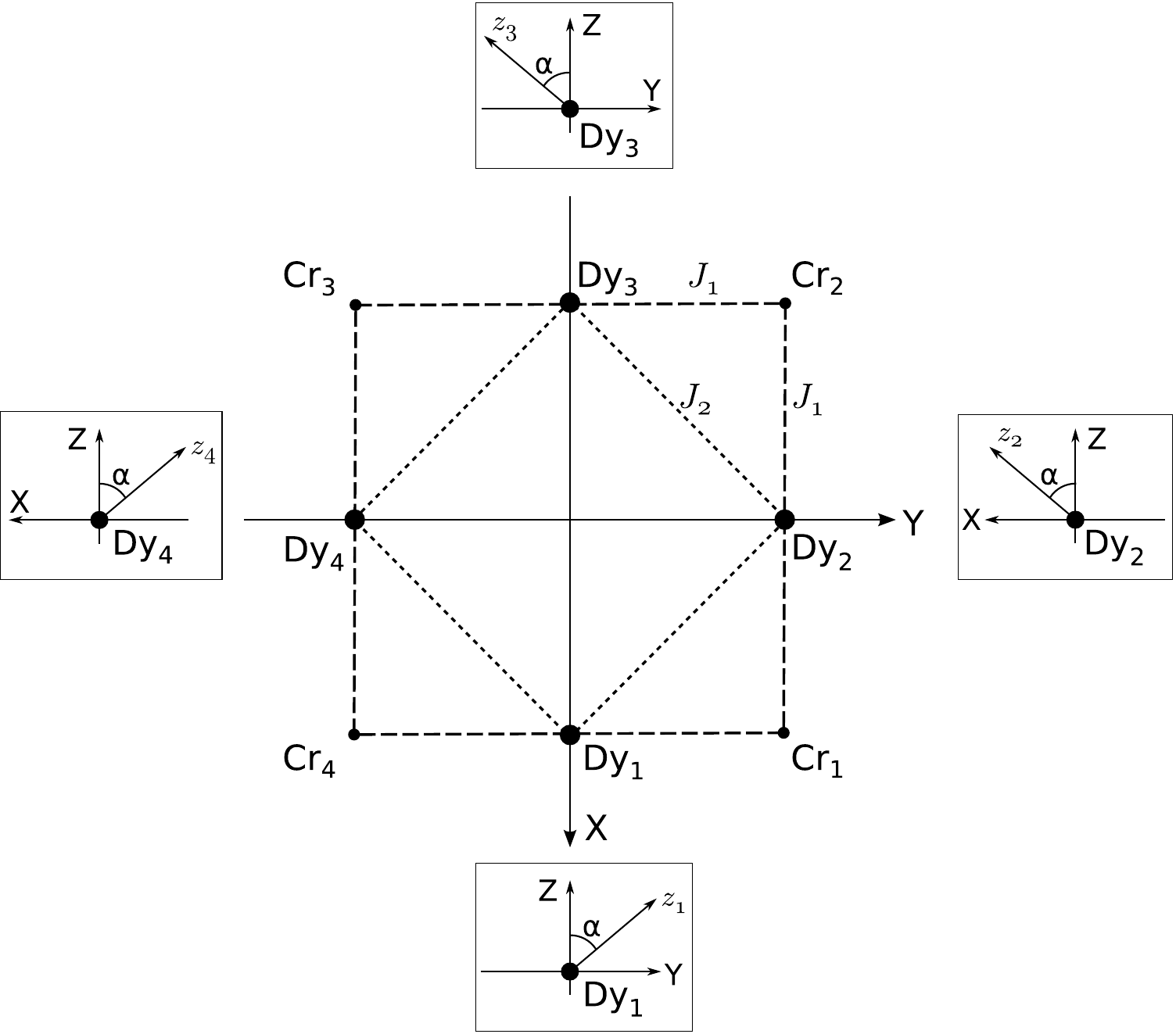} \caption{Schematic
representation of the Dy$_4$Cr$_4$ molecule indicating numbering of atoms and
exchange coupling constants. The boxes show the orientation of the local
anisotropy axes on Dy sites, when viewed from the poles of the $X$ and $Y$
axes. The $Z$ axis points out of the center of the scheme.}
\label{4:fig:ringscheme}
\end{figure}

As a second example we describe in this section the application of the
decorated Ising model to a ring-shaped \ce{Dy4Cr4}
molecule.\cite{AngewChemIntEd_49_7583} Dy$_4$Cr$_4$ consists of alternating
\ce{Dy^{3+}} and \ce{Cr^{3+}} ions forming a closed ring. The four \ce{Dy^{3+}}
ions lie in a plane. The \ce{Cr^{3+}} ions are positioned alternatingly above
and below this plane, ``decorating'' the Dy-Dy bonds. The molecule has $D_{2d}$
symmetry, the \ce{Dy^{3+}} ions lying on $C_2$ axes and the \ce{Cr^{3+}} ions
lying on the mirror planes. We choose a molecular reference frame $XYZ$ so that
$Z$ coincides with the $S_4$ axis and $X$ and $Y$ coincide with the two $C_2$
axes of $D_{2d}$ (Fig.~\ref{4:fig:ringscheme}).

\textit{Ab initio} calculations have been performed in the same way as for the
[DyCuMoCu] chain.\cite{AngewChemIntEd_49_7583} From these, we take again the
$g$-factors of the ground doublet of \ce{Dy^{3+}} and of the isotropic ground
state spin multiplet of \ce{Cr^{3+}} ($d^3$, $S=3/2$):
\begin{equation}\label{4:eq:Dy_gfactors_ring}
\begin{split}
\mathrm{Dy^{3+}}&: \qquad g_z=19.66,\quad g_x=0, \quad g_y=0,\\
\mathrm{Cr^{3+}}&: \qquad g_\mathrm{Cr}=1.97.
\end{split}
\end{equation}
The \ce{Dy^{3+}} Kramers doublet is again very close to the $|M_J=\pm
15/2\rangle$ state, permitting the use of the Ising model. However, the same
calculation predicted the second Kramers doublet at 30\unit{cm^{-1}}, not very
high compared with exchange interaction, which we found in the previous section
$\approx 10\unit{cm^{-1}}$. This should be seen as a warning that our treatment
of the excited Kramers doublets as ``innocent'' may not be entirely correct
here, and thus may lead to discrepancies with experiment. In this respect we
must also note that, for such small excitation energies, the results of the
\textit{ab initio} calculations are not always conclusive on the nature of the
ground state Kramers doublet. In the present case, for instance, another set of
calculations produced a ground state Kramers doublet on \ce{Dy^{3+}} that is
not uniaxial as in Eq.~\eqref{4:eq:Dy_gfactors_ring}, having relatively large
transversal $g$-factors.\cite{AngewChemIntEd_49_7583} The decorated Ising
model would be unusable in this case.  We find however that, assuming the axial
g-factors in Eq.~\eqref{4:eq:Dy_gfactors_ring}, \ce{Dy4Cr4} is an interesting
example of a decorated Ising ring, for which qualitative agreement with
experimental magnetic properties can be obtained. 

Direct ligand bridges connect each \ce{Dy^{3+}} with two neighboring
\ce{Cr^{3+}} ions and two neighboring \ce{Dy^{3+}} ions. Exchange interaction
between these pairs is introduced [Eqs~\eqref{4:eq:exch_DyDy} and
\eqref{4:eq:exch_DyS}]. The Hamiltonian is then given by
Eq.~\eqref{4:eq:HamdecIsing}: $\hat{H}=\hat{h}_1+\hat{h}_2+\hat{h}_3+
\hat{h}_4$, and 
\begin{equation}\label{4:eq:Ham_ring}
\begin{split}
\hat{h}_i(s_i^{z_i},s_{i+1}^{z_{i+1}})&= -J_1 (s^{z_i}_i S_i^{z_i}
+s^{z_{i+1}}_{i+1}
S_i^{z_{i+1}})-J_2s^{z_i}_is^{z_{i+1}}_{i+1}\\
&\quad -\mu_\mathrm{B}( g_\mathrm{Dy}
s^{z_i}_i B^{z_i}+ g_\mathrm{Cr} \mathbf{S}_i \cdot \mathbf{B}),
\end{split}
\end{equation}
where $s^{z_i}_i$ denotes the Ising spin-1/2 variable on Dy$_i$ and $S_i^{z_i}$
denotes the projection of the spin of Cr$_i$ on the magnetic anisotropy axis of
Dy$_i$ (for numbering, see Fig~\ref{4:fig:ringscheme}).  Similarly, $B^{z_i}$
is the projection of the magnetic field on the anisotropy axis of Dy$_i$.
$g_\mathrm{Dy}$ is the $g_z$ factor of \ce{Dy^{3+}}
[Eq.~\eqref{4:eq:Dy_gfactors_ring}].

\begin{figure}\centering
\includegraphics{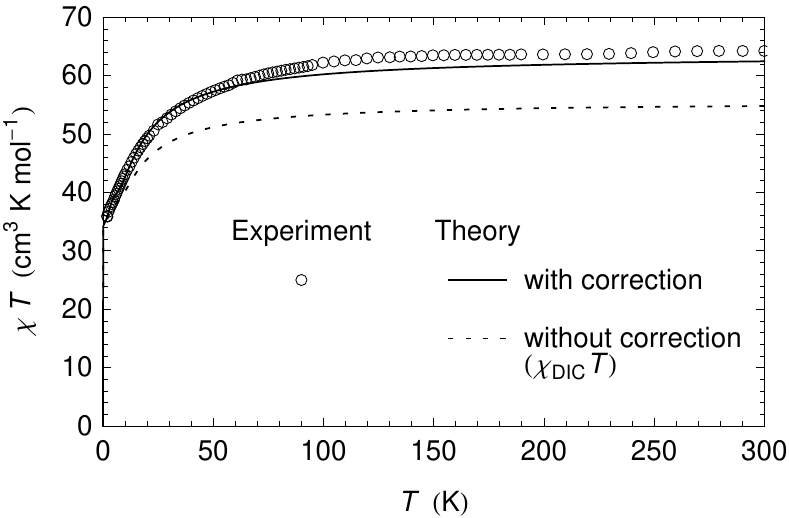}
\includegraphics{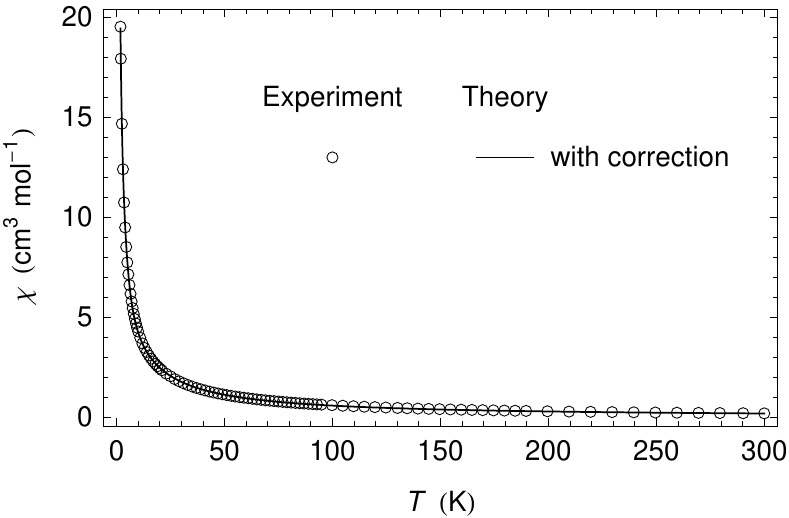}
\caption{Powder magnetic susceptibility of \ce{Dy4Cr4}. The correction on the
theory refers to Eq.~\eqref{4:eq:corrections}.}
\label{4:fig:ringchiTp}
\end{figure}

An interesting difference with the [DyCuMoCu] chain is that here, in
\ce{Dy_4Cr_4}, the four anisotropy axes $z_i$ are not, in general, parallel, a
result of point symmetry instead of translational symmetry. The orientation of
the local anisotropy axis on \ce{Dy^{3+}}, being one of the $g$-tensor main
axes, is restricted by the local $C_2$ symmetry to be either parallel with, or
orthogonal to the local $C_2$ axis. The first possibility can be excluded on
the basis of the experiment; with the $z_i$ pointing radially outwards at each
Dy$_i$, the ground state of the whole molecule is necessarily nonmagnetic,
because the local moments add up to zero, independent of whether the ground
state is F or AF with respect to the Ising spins.  The experimental
susceptibility measurement however indicates a magnetic ground state (nonzero
intercept on the vertical axis in Fig.~\ref{4:fig:ringchiTp}). We must
therefore choose the second case and let the anisotropy axis on each Dy be
orthogonal to the local $C_2$ axis and make an angle of $\alpha$ with the
molecular $Z$-axis (See Fig.~\ref{4:fig:ringscheme}). By applying the symmetry
elements of $D_{2d}$ to one of these anisotropy axes, one obtains the other
three. When $\alpha =0$ the four axes are parallel and point in the same
direction as $Z$. We note that the \textit{ab initio} calculations yielded
$\alpha =37^\circ$. We will need some flexibility in our model however, so we
leave $\alpha$ as a parameter that will be determined from comparison with
experiment. 

In terms of the molecular coordinate system, the projections
on the local anisotropy axes are a function of $\alpha$:
\begin{equation}\label{4:eq:ring_projections}
\begin{split}
S_i^{z_1} &= \cos(\alpha)  S_i^Z + \sin(\alpha) S_i^Y  \\
S_i^{z_2} &= \cos(\alpha) S_i^Z + \sin(\alpha) S_i^X  \\
S_i^{z_3} &= \cos(\alpha) S_i^Z - \sin(\alpha) S_i^Y  \\
S_i^{z_4} &= \cos(\alpha) S_i^Z - \sin(\alpha) S_i^X.
\end{split}
\end{equation}
The same relations hold for the magnetic field, after replacing $S_i$ by $B$.

The fact that only exchange interactions of Ising type appear in
Eq.~\eqref{4:eq:Ham_ring} makes it possible to find analytical solutions of the
eigenvalues and the partition function. From Eqs.~\eqref{4:eq:Ham_ring} and
\eqref{4:eq:ring_projections} we see that the part of $\hat{h}_i$ that involves
$S_i$ is a projection of $\mathbf{S}_i$ on the vector
\begin{equation}\label{4:eq:ring_projectionvector}
-J_1(s_i\hat{\mathbf{z}}_i+s_{i+1}\hat{\mathbf{z}}_{i+1})
-g_\mathrm{Cr}\mu_\mathrm{B}\mathbf{B},
\end{equation}
where $\hat{\mathbf{z}}_i$ is the unit vector along the anisotropy axis of
Dy$_i$ (the superscripts $z_i$ on $s_i$ are left out from now on).
The vector \eqref{4:eq:ring_projectionvector} defines the quantization axis of
$\mathbf{S}_i$, which depends on the states on the neighboring \ce{Dy^{3+}}
sites ($s_i$, $s_{i+1}$). The stronger the coupling ($J_1$) with Dy, the
stronger will be the deviation of the quantization axis from the direction of
$\mathbf{B}$. The eigenvalues of
$\hat{h}_i$ are then 
\begin{equation}\label{4:eq:ring_eigenvalues}
\varepsilon_{iM_S}(s_i,s_{i+1})= b_i M_{S}-J_2 s_i
s_{i+1}-\mu_\mathrm{B} g_\mathrm{Dy} s_i B^{z_i},
\end{equation}
where $M_S= -S,\ldots, S$, and 
\begin{widetext}
\begin{equation}\label{4:eq:ring_bi}
b_i=\sqrt{\frac{J_1^2}{2}(1+4s_is_{i+1}\cos^2\alpha)
+2J_1\mu_\mathrm{B}g_\mathrm{Cr}
(s_i B ^{z_i}+s_{i+1}B^{z_{i+1}})+\mu^2_\mathrm{B} g^2_\mathrm{Cr} B^2}
\end{equation}
\end{widetext}
is the length of the vector in Eq.~\eqref{4:eq:ring_projectionvector}. Some
remarks should be made on the solutions. Eqs.~\eqref{4:eq:ring_bi} and
\eqref{4:eq:ring_projections} (replace $S_i$ by $B$) show that the spectrum in
Eq.~\eqref{4:eq:ring_eigenvalues} is not the same for every unit $i$, as it was
in the [DyCuMoCu] chain, unless $\mathbf{B}$ is applied along the $Z$ axis.
This means that also the transfer matrices $T_i$ will be different and that we
have to use Eq.~\eqref{4:eq:partfunc_transmatr} instead of
Eq.~\eqref{4:eq:partfunc_periodic} for the partition function. A second remark
concerns the quantum number $M_S$. The lowest energy in
Eq.~\eqref{4:eq:ring_eigenvalues} is always given by $M_S=-S$, but note that
the axis to which this quantization refers is not invariant; in particular, it
changes with strength and direction of applied field, so that $M_S$ does not
represent a real conserved quantity that could be responsible for level
crossings of the ``internal'' type. Such crossings do not occur in \ce{Dy4Cr4}.

We conclude the solution by finding the partition function $\mathcal{Z}$.
Substituting Eq.~\eqref{4:eq:ring_eigenvalues} in
Eq.~\eqref{4:eq:subpartitionfunction} we find
\begin{align*}
\Psi_i(s_i,s_{i+1})&=\sum_{M_S=-S}^Se^{-\beta \varepsilon_{iM_S}}\\
&=\frac{\sinh [\beta b_i(2S+1)/2]}{\sinh [\beta b_i/2]}\\
&\quad \times \exp[\beta(J_2 s_is_{i+1}
+\mu_\mathrm{B} g_\mathrm{Dy} s_i B^{z_i})]
\end{align*}
With $T_i$ as defined in Eq.~\eqref{4:eq:Tmatrix}, we obtain the partition
function
\begin{equation*}
\mathcal{Z}=\trace (T_1T_2T_3T_4).
\end{equation*}

Let us now compare the theoretical results with experiment. A great amount of
information on the values of the parameters $\alpha$, $J_1$ and $J_2$ can be
obtained by inspection of the powder $\chi T$ curve
(Fig.~\ref{4:fig:ringchiTp}). The nonzero intercept $\chi T|_{T\to
0}=34.6\unit{cm^3\,K\,mol^{-1}}$ indicates a magnetic ground
state.\cite{kaHn_mol_magn} Now from the general theory we know that the ground
state is either F ($\uparrow\uparrow\uparrow\uparrow$) or AF
($\uparrow\downarrow\uparrow\downarrow$) with respect to the \ce{Dy^{3+}}
spins;\footnote{This result, which was found in Section \ref{4:sec:eigenstates}
for a periodic ring or chain, is valid here because we are considering the
eigenstates of \ce{Dy4Cr4} in the \emph{absence} of magnetic field, in which
case the ring is effectively cyclic symmetric.} AF is nonmagnetic so we decide
that the ground state must be F. 

Incidentally, we can precisely delineate the regions in parameter space where
the ground state is F or AF:
\begin{subequations}\label{4:eq:ring_groundstates}
\begin{align}
\mathrm{F}\;(\uparrow\uparrow\uparrow\uparrow)&:
J_2
>-\sqrt{2}S\left(\sqrt{1+\cos^2\alpha}-\sin\alpha\right)|J_1|,
\label{4:eq:ring_groundstates_F}\\ 
\mathrm{AF} (\uparrow\downarrow\uparrow\downarrow)&: 
J_2 < -\sqrt{2}S\left(\sqrt{1+\cos^2\alpha}-\sin\alpha\right)|J_1|.
\label{4:eq:ring_groundstates_AF}
\end{align}
\end{subequations}

A second piece of information comes from the increase of $\chi T$ with
increasing temperature. This is partly but not completely due to the occupation
of excited Kramers doublets, as one can show by subtracting the contribution
of the latter, obtained from the \textit{ab initio} calculations (not shown
here). There must still be an antiferromagnetic interaction to explain the
increase. Since the \ce{Dy^{3+}} are already known to be ferromagnetically
aligned, the only possibility is that the \ce{Cr^{3+}} spins couple
antiferromagnetically with \ce{Dy^{3+}}, or $J_1<0$.

With this information, we can determine the angle $\alpha$.  At 0\unit{K},
$\chi T$ is determined by the magnetic moment in the ground state
only.\cite{grIffith_TTMI,kaHn_mol_magn} In the F
($\uparrow\uparrow\uparrow\uparrow$) state,
$\chi_{XX}T|_{T\rightarrow0}=\chi_{YY}T|_{T\rightarrow 0}=0$ by symmetry and
\begin{equation}\label{4:eq:chizzT0} 
\chi_{ZZ}T|_{T\to
0}=\frac{N_A\mu_\mathrm{B}^2}{k} \Bigl\lvert\Bigl\langle F
\Big|g_\mathrm{Dy}\cos\alpha
\sum_{i=1}^4s_i+g_\mathrm{Cr}\sum_{i=1}^4S_i^Z\Big|F\Bigr\rangle\Bigr\rvert^2,
\end{equation} 
so $\chi T|_{T\to 0}=\frac{1}{3}\chi_{ZZ} T|_{T\to 0}$.
With the help of
Eqs.~\eqref{4:eq:ring_projections}--\eqref{4:eq:ring_bi} and the fact that, in
the ground state, $M_S=-S$ in Eq.~\eqref{4:eq:ring_eigenvalues}, we can
evaluate Eq.~\eqref{4:eq:chizzT0} to find
\begin{equation*}
\begin{split}
\chi T|_{T\rightarrow 0} &= \frac{N_\mathrm{A} \mu_\mathrm{B}^2}{3k}
4\cos^2\alpha\biggl(g_\mathrm{Dy}\\
&\quad +\sign (J_1) 2\sqrt{2}g_\mathrm{Cr}S
\frac{1}{\sqrt{1+\cos^2\alpha}}\biggr)^2.
\end{split}
\end{equation*}
This is a strictly decreasing function of $\alpha$ that can be used to derive
$\alpha$ from the experimental value 
$\chi T|_{T\rightarrow 0}=34.6\unit{cm^3\,K\,mol^{-1}}$, and the knowledge
that $\sign(J_1)=-1$. This gives $\alpha=49^\circ$. 

\begin{figure}\centering
\includegraphics{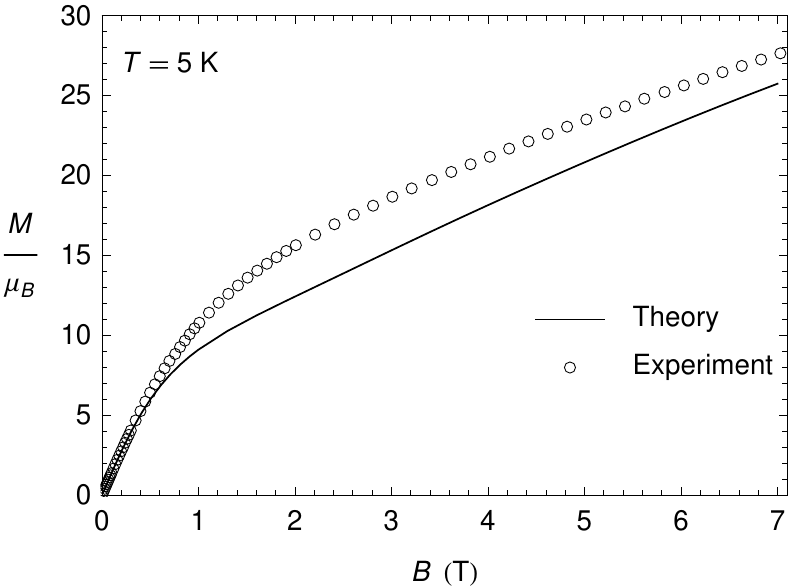}
\caption{Powder magnetization of \ce{Dy4Cr4}. A linear corrections of
0.5\unit{\mu_B/T} has been added to the theoretical curve, according to
Eq.~\eqref{4:eq:corrections_M}.}
\label{4:fig:ringMp}
\end{figure}

We derive values for $J_1$, $J_2$, and $\alpha$ by a least-squares fitting of
$\chi$. As before, a correction for the contribution of excited \ce{Dy^{3+}}
Kramers doublets is provided by the \textit{ab initio} calculations and
applied following Eq.~\eqref{4:eq:corrections}.
 The fitting yields
\begin{equation*}
J_1=-8.9\unit{cm^{-1}},\quad J_2=5.2\unit{cm^{-1}},\quad \alpha=49.2^\circ.
\end{equation*}
The comparison with experiment is shown in Figs.~\ref{4:fig:ringchiTp} and
\ref{4:fig:ringMp}. Note that the magnetic properties are reported per mole
($\chi$ and $\chi T$) or per molecule ($M$) of \ce{Dy4Cr4} and not per DyCr
unit. Note also that $\sign(J_1)=-1$, that $J_2$ satisfies
Eq.~\eqref{4:eq:ring_groundstates_F}, and that $\alpha$ agrees with the value
derived above. 

The agreement of magnetization curves (Fig.~\ref{4:fig:ringMp}) is not as good
as it was for the [DyCuMoCu]$_\infty$ chain, although the qualitative
properties seem to correspond. In particular, we mention the strong linear
increase of $M$ at higher fields ($\approx 2.6\unit{\mu_B/T}$), which is due to
the gradual orientation of the \ce{Cr^{3+}} spins to the magnetic field [see
discussion connected with Eq.~\eqref{4:eq:ring_projectionvector}], and, to a
smaller extent, also to the correction of 0.5\unit{\mu_B/T}, a non-negligible
linear contribution to magnetization, which is due to the low-lying excited
Kramers doublets.  

As was mentioned before, the discrepancies are not unexpected given a
low-lying first excited Kramers doublet of \ce{Dy^{3+}}, which could undermine
the assumptions underlying the decorated Ising model. Note also that we could
not take the \textit{ab initio} value of 37$^\circ$ for $\alpha$. Leaving
$\alpha$ as a parameter can be seen as a partial compensation for the
inaccuracies of the model and the \textit{ab initio} results. 

\section{Conclusion}

We have shown that the decorated Ising model is a valid model for the magnetic
properties of certain lanthanide-containing magnetic compounds, if the crystal
field spectrum of the lanthanide ion satisfies certain properties. The most
important of these is the requirement of a ground state Kramers doublet with
completely uniaxial magnetic anisotropy (this statement is simplified, see
Section \ref{4:sec:Dy} for the correct details).  It is a remarkable fact that
precisely this property has been established by multiconfigurational \textit{ab
initio} calculations on several \ce{Dy^{3+}} centers that are part of
polynuclear molecular magnets. Perhaps the best known example is the \ce{Dy3}
triangle, where the Ising properties of \ce{Dy^{3+}} were used to explain the
nature of the ground state.\cite{AngewChemIntEd_47_4126}

We have focused on \ce{Dy^{3+}} as lanthanide ion because this is a much-used
lanthanide in current synthetic research in molecular magnetism (witness both
compounds in this paper) and because computational results showing that it
meets the requirements for an Ising spin are available. However, there is no
reason to assume that the findings are unique to \ce{Dy^{3+}}. We expect that
other lanthanides with high momentum (e.g., \ce{Er^{3+}}) will exhibit the same
uniaxial anisotropy in certain ligand environments and that examples of
decorated Ising chains based on lanthanides other than \ce{Dy^{3+}} will be
found in the future.

\begin{acknowledgments}
We thank Liviu Ungur for providing results of the \textit{ab initio}
calculations. We thank the referee for useful suggestions and comments.
W.~V.d.H. acknowledges financial support from the Research Foundation - Flanders (FWO).
\end{acknowledgments}

\end{document}